%% file: main.tex
\documentclass[10pt,aps,prc,floatfix,twocolumn,superscriptaddress,nofootinbib]{revtex4-2}

\usepackage[labelfont={small},subrefformat=parens,caption=false]{subfig}

\usepackage[dvipsnames]{xcolor}
\definecolor{RED}{named}{red}
\usepackage{amsfonts,amsmath,amssymb,bm}
\usepackage{graphicx}
\usepackage{placeins}
\usepackage{braket}

\usepackage{dcolumn}
\newcolumntype{d}[1]{D{.}{.}{#1}}
\usepackage{cellspace}
\usepackage{xspace}
\usepackage{isotope}
\usepackage{xparse}
\usepackage{physics}
\usepackage[normalem]{ulem}

\usepackage{microtype} 

\usepackage[pdfpagelabels, pdfencoding=auto, psdextra]{hyperref}
\hypersetup{%
 pdfsubject={Many-body perturbation theory for the nuclear equation of state up to fifth order},
 pdfkeywords={nuclear physics, nuclear theory, neutron stars, chiral effective field theory, nuclear matter, equation of state, EOS, many-body perturbation theory, MBPT, Monte Carlo integration, VEGAS},
 unicode = true,
 breaklinks = true,
 colorlinks = true,
 linkcolor = blue,
 menucolor = blue,
 citecolor = blue,
 urlcolor = blue
}

\definecolor{bobcatgreen}{rgb}{0.3,0.6,0.1}
\definecolor{scarlet}{rgb}{0.73,0,0}

\graphicspath{{./figures/}}

\newcommand{\orcid}[1]{\href{https://orcid.org/#1}{\includegraphics[scale=0.055]{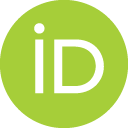}}}
\input{buqeye_macros}

\input{macros}

\begin{document}

\title{Many-body perturbation theory for the nuclear equation of state up to fifth order}

\author{C.~Drischler~\orcid{0000-0003-1534-6285}}
\email{drischler@ohio.edu}
\affiliation{Department of Physics and Astronomy, \href{https://ror.org/01jr3y717}{Ohio University}, Athens, Ohio~45701, USA}
\affiliation{\href{https://ror.org/03r4g9w46}{Facility for Rare Isotope Beams}, \href{https://ror.org/05hs6h993}{Michigan State University}, East Lansing, Michigan~48824, USA}

\author{K.~S.~McElvain~\orcid{0000-0002-1405-7935}}
\email{kenmcelvain@berkeley.edu}
\affiliation{Department of Physics, \href{https://ror.org/01an7q238}{University of California, Berkeley}, California~94720, USA}
\affiliation{Nuclear Science Division, \href{https://ror.org/02jbv0t02}{Lawrence Berkeley National Laboratory},
Berkeley, California~94720, USA}

\author{P.~Arthuis~\orcid{0000-0002-7073-9340}}
\email{pierre.arthuis@ijclab.in2p3.fr}
\affiliation{\href{https://ror.org/03xjwb503}{Universit{\'e} Paris-Saclay}, \href{https://ror.org/03gc1p724}{CNRS/IN2P3, IJCLab}, 91405 Orsay, France}

\date{\today}

\begin{abstract} 

We present an automated, GPU-accelerated framework for many-body perturbation theory (MBPT) calculations of the zero-temperature nuclear equation of state (EOS) based on chiral nucleon-nucleon (NN) and three-nucleon (3N) interactions. Automated diagram generation and evaluation enable the computation of all diagrams up to fifth order in the MBPT expansion at the normal-ordered two-body level in infinite matter, with residual three-body contributions explicitly included up to third order. Multi-GPU acceleration of 3N normal ordering, a novel Monte Carlo integrator (called \texttt{PVegas}), and further advances in high-performance computing enable us to evaluate all 840 fifth-order diagrams with controlled numerical uncertainties. We investigate the MBPT convergence up to fifth order in pure neutron matter (PNM) and symmetric nuclear matter (SNM) for two sets of chiral interactions, study neutron star matter, and present fourth-order results for asymmetric matter including normal-ordered 3N forces. The framework enables systematic MBPT studies with harder interactions and benchmarks against nonperturbative methods. It can be further extended to finite-temperature EOS calculations and to improved uncertainty quantification using emulation and resummation techniques.

\end{abstract}

\maketitle

\section{Introduction}
\label{sec:intro}

Substantial progress has been made in microscopic predictions of the nuclear equation of state (EOS) in recent years, with important consequences for neutron-rich nuclei, heavy-ion collisions, and neutron stars~\cite{Drischler:2021kxf,Sorensen:2023zkk,MUSES:2023hyz,Chatziioannou:2024jsr,Agarwal:2025ezo}.
Chiral effective field theory (EFT)~\cite{Epelbaum:2008ga,Machleidt:2011zz,Hammer:2019poc,Epelbaum:2019kcf} has become a powerful framework for deriving microscopic nuclear interactions consistent with the symmetries of low-energy quantum chromodynamics (QCD).
Furthermore, advances in many-body theory, such as the development of the in-medium similarity renormalization group (IMSRG) for infinite matter~\cite{Zhen:2025gfy}, 
quantum Monte Carlo (QMC) methods for neutron-rich matter~\cite{Lonardoni:2019ypg,Arthuis:2022ixv,Fore:2024exa,Lovato:2026erx},\footnote{%
\rev{For other recent developments in QMC-based calculations of nuclear systems, see Refs.~\cite{Hu:2026cmr,Hu:2026ilr,Hu:2026ysy}.}%
} 
diagrammatic Monte Carlo for nuclear pairing~\cite{Brolli:2025ehf}, 
\rev{coupled-cluster (CC) theory~\cite{Hagen:2013yba,Baardsen:2013vwa,Jiang:2022tzf},} 
and Gorkov self-consistent Green's function (SCGF) for infinite matter~\cite{FrancescoMarino:2026dzw}, as well as fast and accurate emulators~\cite{Duguet:2023wuh,Melendez:2022kid,Drischler:2022ipa,Cook:2024toj,Frame:2017fah}, have enabled increasingly accurate studies of nuclear matter with theoretical uncertainties rigorously quantified (e.g., see Refs.~\cite{Curry:2025pna,Armstrong:2025tza,Somasundaram:2024zse,Jiang:2022oba} for recent applications of EOS emulators).

At the same time, many-body perturbation theory (MBPT)~\cite{Tichai:2020dna,Drischler:2021kxf} has experienced a renaissance as a computationally inexpensive, yet accurate, approach to solving the nuclear many-body Schr{\"o}dinger equation perturbatively (see also Refs.~\cite{Tichai:2021ewr,Arthuis:2018yoo,Arthuis:2020tjz}).
MBPT provides a versatile framework for constraining the nuclear EOS as a function of the density, isospin asymmetry, and temperature~\cite{Drischler:2017wtt,Wellenhofer:2018dwh,Wellenhofer:2021eis,Keller:2020qhx,Wen:2020nqs,Wen:2024shw,Keller:2022crb}, especially when used with soft interactions obtained from, e.g., the similarity renormalization group (SRG)~\cite{Hebeler:2010xb,Huther:2019ont,Arthuis:2024mnl}.
This framework also allows for estimating many-body uncertainties due to truncating the MBPT expansion at a finite order~\cite{Svensson:2025jde} and for assessing common many-body approximations~\cite{Yang:2021vxa,Rothman:2025rua}, such as neglecting residual 3N contributions.

In many of the cited works, MBPT calculations based on chiral nucleon-nucleon (NN) and three-nucleon (3N) forces have been performed up to second or third order in the many-body expansion, which already capture important correlation effects beyond the Hartree-Fock approximation. 
In Refs.~\cite{Drischler:2017wtt,Leonhardt:2019fua}, the first fourth-order MBPT calculations for infinite matter were carried out, although with 3N contributions only included up to third order at the normal-ordered two-body level~\cite{Hebeler:2020ocj}.
While these studies have provided valuable insights into infinite matter up to twice the nuclear saturation density, questions remain about the MBPT convergence\footnote{%
Although the MBPT expansion is an asymptotic series, we use the term ``convergence'' throughout the paper in a practical sense to describe the improvement in accuracy with increasing MBPT order, up to the onset of divergence.
} and about the significance of neglected higher-order contributions in obtaining more accurate EOS predictions.
These questions motivate explicit higher-order MBPT calculations.

Systematically extending MBPT calculations to higher orders also improves the estimation of MBPT truncation errors~\cite{Svensson:2025jde} and more stringent benchmarks against nonperturbative methods with modern nuclear interactions~\cite{Tews:2012fj,Drischler:2016djf,Lovato:2022apd}.
Furthermore, higher-order MBPT calculations can be used to guide resummation techniques, such as Pad{\'e} approximations and reduced basis methods~\cite{Demol:2019yjt,Demol:2020mzd,Tichai:2020dna}, for extracting converged results from slowly converging MBPT expansions.
However, EOS calculations beyond third order that treat NN and 3N contributions on an equal footing have so far been infeasible due to the rapid growth of MBPT diagrams and the substantial computational cost of including 3N contributions at both the normal-ordered two-body and residual three-body levels~\cite{Drischler:2021kxf}.

In this work, we present an automated, GPU-accelerated framework for MBPT calculations of the zero-temperature EOS with chiral NN and 3N interactions. 
By combining automated diagram generation~\cite{Arthuis:2018yoo} with automated diagram evaluation~\cite{Drischler:2017wtt}, we derive all MBPT diagrams up to sixth order and evaluate them up to fifth order at the normal-ordered two-body level (``MBPT(5)''),\footnote{%
\rev{For related advances in high-order MBPT calculations of ground-state energies and radii of closed-shell nuclei, see Ref.~\cite{Li:2026njw}.}%
} 
including residual 3N contributions up to third order.
Advances in high-performance computing, such as multi-GPU acceleration of 3N normal ordering and a novel Monte Carlo integrator with importance sampling (called \texttt{PVegas}), allow us to evaluate all 840 fifth-order diagrams with controlled numerical uncertainties. 
For comparison, there are only 3 and 39 MBPT diagrams at third and fourth order, respectively.
Using this framework, we investigate the MBPT convergence in pure neutron matter (PNM) and symmetric nuclear matter (SNM), study neutron-star matter, and present parametric models that describe our explicit asymmetric-matter calculations, thereby significantly extending and improving previous MBPT calculations.

The remainder of this work is organized as follows.
In Sec.~\ref{sec:mbpt_diagrams}, we briefly summarize the MBPT expressions at the NN and 3N levels considered here. 
Explicit MBPT expressions are given up to third order; a complete list of the MBPT expressions up to sixth order can be found in the accompanying GitHub repository~\cite{Drischler_Cheat_sheets}.
Section~\ref{sec:mbpt_eval} then describes our automated MBPT framework to evaluate these MBPT expressions efficiently and with controlled uncertainties. 
\rev{Our main results based on two sets of nuclear interactions are presented in Secs.~\ref{sec:results} to~\ref{sec:fit_interactions}, including a convergence study of MBPT up to fifth order, constraints on low-density EOS parameters, and explicit asymmetric matter calculations.}
This article concludes in Sec.~\ref{sec:summary_outlook} with a summary and an outlook of future high-order MBPT calculations of nuclear matter.
Three Appendixes provide additional information.
While Appendix~\ref{app:add_results} contains supplemental results, Appendixes~\ref{app:traces} and~\ref{app:comp_scaling} discuss the optimized evaluation of spin-isospin traces and the computational scaling of our MBPT framework, respectively.
We use natural units in which $\hbar = c = 1$. 

\section{Many-body expressions} \label{sec:mbpt_diagrams}

We briefly describe in this section our MBPT calculations in nuclear matter with chiral nucleon-nucleon (NN) and three-nucleon (3N) interactions, and refer the reader to Ref.~\cite{Drischler:2021kxf} for a comprehensive review article and Ref.~\cite{Drischler_Cheat_sheets} for a machine-readable compilation of all many-body expressions used in this work.
The MBPT contributions to the energy density up to third order at the two-body level are well-known in the literature (see, e.g., Refs.~\cite{STEVENSON03,shavitt2009many,Holt:2016pjb,Arthuis:2018yoo,Arthuis:2020tjz}).
Expressed in the single-particle basis $\ket{p} = \ket{\vb{k}_p\sigma_p\tau_p}$ with momentum vector $\vb{k}_p$, spin projection $\sigma_p$, and isospin projection $\tau_p$, they are given by\footnote{%
We use the labels $\{a,b,c,\ldots \}$ for particles and $\{i,j,k,\ldots \}$ for holes, with ``$p$'' being a generic label.}
\begin{align}
\frac{{E}^{(\text{HF})}}{V} &= \sum_i \eps[i] - \frac{1}{2} \sum \limits_{\substack{ij}} \mel{ij}{\Vmed{(\xi=2/3)}}{ij} \,, \label{eq:E_HF}\\
\frac{{E}^{(2)}}{V} &= 
\sum \limits_{\substack{ij\\ab}} \frac{\mel{ij}{\Vmed{}}{ab} \mel{ab}{\Vmed{}}{ij}}{4D_{ijab}} \,, \label{eq:E_2nd}\\
\frac{E^{(3)}_1}{V} &= 
\sum \limits_{\substack{ijkl\\ab}} \frac{\mel{ij}{\Vmed{}}{ab} \mel{kl}{\Vmed{}}{ij} \mel{ab}{\Vmed{}}{kl}}{8D_{ijab}D_{klab}}\,, \label{eq:E_3rd_hh} \\
\frac{{E}^{(3)}_2}{V} &= \sum \limits_{\substack{ijk\\abc}} \frac{\mel{ij}{\Vmed{}}{ab} \mel{ak}{\Vmed{}}{ic} \mel{bc}{\Vmed{}}{jk}}{{D_{ijab}D_{jkbc}}}\,, \label{eq:E_3rd_ph} \\
\frac{{E}^{(3)}_3}{V} &= 
\sum \limits_{\substack{ij\\abcd}} \frac{\mel{ij}{\Vmed{}}{ab} \mel{ab}{\Vmed{}}{cd} \mel{cd}{\Vmed{}}{ij}}{8{D_{ijab}D_{ijcd}}}\,. \label{eq:E_3rd_pp} 
\end{align}
Here, we have used the shorthand notation for the respective hole and particle summations,
\begin{equation}
\sum_p \rightarrow \sum \limits_{\sigma_p\tau_p} \int \frac{\dd{\vb{k}}_p}{(2\pi)^3}
\begin{cases}
\nk[p] & \text{for $p \in \{i,j,k,\ldots \}$}\,,\\
1 - \nk[p] & \text{for $p \in \{a, b, c,\ldots \}$}\,.
\end{cases}
\end{equation}
In the zero-temperature limit, the Fermi-Dirac distribution function is equal to the Heaviside step function
\begin{equation}
    \nk = \theta(\kF[\tau] - k )\,, \quad \text{where} \quad \kF[\tau] =\sqrt[3]{3\pi^2 \iso{n}{\tau}} 
\end{equation}
is the Fermi momentum corresponding to the nucleon density of species $\tau \in \{ n,p\}$, $\iso{n}{\tau}$.
The terms in the energy denominators of the MBPT expressions read
\begin{equation} \label{eq:Dijkabc}
\begin{split}
D_{ijk\ldots abc\ldots} &= \eps[i] + \eps[j] + \eps[k] + \ldots \\
&\quad -\eps[a] - \eps[b] - \eps[c] - \ldots\,,
\end{split}
\end{equation}
with the (isospin-dependent) single particle energy in a Hartree-Fock (HF) spectrum
\begin{equation} \label{eq:spe}
 \eps[i] = \frac{k_i^2}{2m_n^{(\tau_i)}} + \sum \limits_{\substack{j}} \mel{ij}{\Vmed{(\xi=1/2)}}{ij} \,,
\end{equation}
where $m_n^{(\tau)}$ denotes the nucleon mass and the second term on the right side corresponds to the nucleon self-energy, $\Sigma^\text{(HF)}$. 
Our calculations do not apply the isospin averaging approximation (e.g., see Ref.~\cite{Drischler:2015eba}) and instead account for the isospin dependence of Eq.~\eqref{eq:spe} explicitly.

Following the standard approach~\cite{Hebeler:2020ocj,Drischler:2021kxf,Rothman:2025rua}, we use normal ordering to include important 3N contributions at the (normal-ordered) two-body level in our calculations. 
The nuclear interaction $\Vmed{}$ in Eqs.~\eqref{eq:E_HF}--\eqref{eq:E_3rd_pp} and~\eqref{eq:spe} is then given by the sum of the (antisymmetrized) free-space NN interaction and a density-dependent effective two-body potential obtained by summing one of the three interacting nucleons over the occupied states of the HF reference state,
\begin{multline} \label{eq:Veff}
        \mel{p_{2'}p_{3'}}{\Vmed{(\xi)}}{p_2p_3} = \mel{p_{2'}p_{3'}}{\mathcal{A}_{12}V_\text{NN}}{p_{2}p_{3}} \\
        + \xi \sum \limits_{i} \mel{ip_{2'}p_{3'}}{\mathcal{A}_{123}V_\text{3N}}{ip_{2}p_{3}}\,.
\end{multline}
The combinatorial factor $\xi$ is determined by Wick's theorem and depends on the many-body calculation of interest.
For brevity, we omit the superscript specifying its value wherever $\xi=1$; that is, beyond the HF level.
The two- and three-body antisymmetrizer is denoted by $\mathcal{A}_{12}$ and $\mathcal{A}_{123}$, respectively.

In addition to Eqs.~\eqref{eq:E_HF}--\eqref{eq:E_3rd_pp}, MBPT up to third order (with NN and 3N forces) includes (effectively) 15 residual 3N terms, which involve 3N interactions explicitly and thus cannot be calculated in terms of the effective two-body interaction~\eqref{eq:Veff} only. 
Hence, they are usually neglected based on the expectation that they are small compared with the normal-ordered two-body contributions at this order~\cite{Rothman:2025rua}.
To our knowledge, in infinite matter, this expectation has only been benchmarked at second order by explicit calculations of Eq.~\eqref{eq:E_2nd_res} with several chiral interactions~\cite{Drischler:2019xuo,Alp2025}.
Here, we include all 15 diagrams to shed light on the residual 3N contributions up to third order in infinite matter.
At second order, the residual 3N term is
\begin{equation} \label{eq:E_2nd_res}
    \frac{{E}_\mathrm{res}^{(2)}}{V} = 
    \sum \limits_{\substack{ijk\\abc}} \frac{\mel{ijk}{\mathcal{A}_{123}V_\mathrm{3N}}{abc}\mel{abc}{\mathcal{A}_{123}V_\mathrm{3N}}{ijk}}{36D_{ijkabc}}\,.
\end{equation}
At third order, the remaining 14 residual 3N terms were derived by Hu~\etal~\cite{Hu:2018dza}, who also studied their importance in closed-shell nuclei from \isotope[4]{He} to \isotope[48]{Ca}.
Their findings (in those atomic nuclei) suggest that the dominant 3N contributions to the ground state energy are due to the effective two-body potential~\eqref{eq:Veff}, as expected.
We rederived the analytic expressions of these diagrams and provide a complete list in the companion GitHub repository~\cite{Drischler_Cheat_sheets}.

\begin{table}[tb]
\renewcommand{\arraystretch}{1.1}
\caption{%
Number of diagrams $N_\text{diag}$ at order $n$ in MBPT up to fifth (third) order at the normal-ordered two-body (three-body) level. 
In practice, $N_\text{diag}^\text{(eff)} \leqslant N_\text{diag}$ diagrams need to be evaluated due to the presence of anomalous and complex-conjugated pairs of diagrams.
The quoted dimensions of the integrals $N_\text{dim}$ account for momentum conservation and can be further lowered by three by exploiting the rotational invariance of the integrand functions (see Sec.~\ref{sec:mbpt_eval}).  
Two-body diagrams are evaluated in $N_\text{grp}$ groups that have the same particle-hole composition.
Three-body diagrams are individually evaluated, and thus $N_\text{grp}$ is equal to the number of diagrams in these cases.
The expressions of the diagrams can be found in Ref.~\cite{Drischler_Cheat_sheets}. %
}
\label{tab:mbpt}
\begin{ruledtabular}
\begin{tabular}{llllll}
level & MBPT$(n)$ & $N_\text{diag}$\footnotemark[1] & $N_\text{diag}^\text{(eff)}$ & $N_\text{dim}$ & $N_\text{grp}$\\ \hline
two-body     & HF    & 1           & 1   &    6   & 1     \\
      & 2     & 1           & 1   &    9  & 1     \\
      & 3     & 3           & 3  &   18   & 2      \\
      & 4     & 39          & 24  &   24   & 4     \\
      & 5     & 840            & 375    & 30   & 5         \\
      & 6\footnotemark[2]     &  27300           & 11269   & 36  & 6         \\
three-body     & 2     & 1           &  1   &   18    & 1    \\
      & 3     & 19          &  14\footnotemark[3]   &  21; 24; 27 & 14
\end{tabular}
\end{ruledtabular}
\footnotetext[1]{At the two-body level, the sequence $N_\text{diag}(n)$ is known as the integer sequence A064732. See also Ref.~\cite[Ch.~5]{STEVENSON03}.} 
\footnotetext[2]{We have implemented these diagrams, but due to the high computational costs of evaluating them, we do not consider them further in this paper.}
\footnotetext[3]{See Fig.~1 in Ref.~\cite{Hu:2018dza} for the MBPT diagrams.}
\end{table}

Beyond third order, the number of MBPT diagrams $N_\text{diag}$ increases rapidly, as shown in Table~\ref{tab:mbpt}.
For example, at the normal-ordered two-body level, there are 39 and 840 diagrams at the fourth and fifth orders, respectively. 
Although in practice the presence of anomalous and complex-conjugated pairs of diagrams helps reduce the number of diagrams to be computed, $N_\text{diag}^\text{(eff)} \leqslant N_\text{diag}$, there are still, respectively, $N_\text{diag}^\text{(eff)} = 24$ and $375$ diagrams left to be evaluated at these orders.

While the diagrams summarized in Table~\ref{tab:mbpt} can (in principle) be derived manually, an automated approach~\cite{PALDUS19731,WONG19739,KALDOR1976432,CSEPES19881,LYONS199491,STEVENSON03,Li2023PhD} is both more efficient and less error-prone.
Arthuis~\etal~\cite{Arthuis:2018yoo,Arthuis:2020tjz,Tichai:2021ewr} have developed an automated diagram generator (ADG) capable of generating the diagrams and associated analytic expressions in several many-body formalisms, including MBPT, up to arbitrary orders.
We use (and extend) their publicly available ADG to derive the MBPT expressions up to sixth order at the (normal-ordered) two-body level in a tabular format~\cite{Drischler_Cheat_sheets}. 
The expressions match our independent derivation up to fourth order in Ref.~\cite{Drischler:2017wtt}.
Omitted higher-order expressions could be generated accordingly, but the 27300 and 1232280 diagrams already present at sixth and seventh order, respectively, pose a formidable computational challenge \rev{due to both their sheer number and the increasing complexity of individual diagrams. 
Instead of such explicit high-order MBPT calculations, a nonperturbative many-body approach may therefore be more practical.
For example, CC theory with singles and doubles excitations (CCSD), the IMSRG truncated at the normal-ordered two-body level, and the SCGF with the algebraic diagrammatic construction at order 3 [ADC(3)] already resum important classes of diagrams to all orders, including ladder diagrams and chains of particle-hole diagrams~\cite{Hagen:2013nca,Hergert:2015awm,Soma:2020xhv}. 
However, unlike our MBPT(5) calculations, these nonperturbative calculations are not diagrammatically complete up to fifth order, as discussed in detail, e.g., in Ref.~\cite{Li:2026njw} in the context of nuclear structure calculations.}

In this work, we consider all the many-body diagrams summarized in Table~\ref{tab:mbpt} up to fifth order, unless stated otherwise.
The sum of all diagrams gives the total energy density, $E_\text{tot}/V$.
We have also implemented all NN diagrams at sixth order and performed proof-of-principle evaluations with low statistics only.
However, due to the high computational cost of evaluating them at present, exceeding our resources, we do not consider them further here.
That means, in our calculations, all MBPT diagrams are completely accounted for up to third order at the two- and three-body level, while only the residual 3N terms are neglected at the fourth and fifth order.
We provide a comprehensive compilation of the MBPT expressions corresponding to Table~\ref{tab:mbpt}, including \LaTeX\ code and the sixth-order diagrams, in the companion GitHub repository~\cite{Drischler_Cheat_sheets}, enabling practitioners to use and extend them in their research. 
Our improved MC framework for MBPT allows for the systematic calculation of these many-body diagrams, as discussed in the next section.

\section{Automated evaluation of many-body expressions} 
\label{sec:mbpt_eval}

In this section, we discuss our computational framework for the automated evaluation of the many-body expressions described in Sec.~\ref{sec:mbpt_diagrams}.
The framework is a significantly extended and improved implementation of the MC framework introduced in Ref.~\cite{Drischler:2017wtt}, which was more recently applied to finite temperatures~\cite{Keller:2020qhx,Keller:2022crb}.\footnote{%
\rev{Early applications of the SCGF method to finite-temperature calculations can be found in Ref.~\cite{Rios:2008fz}, while more recent developments are discussed in Refs.~\cite{Rios:2020oad,Carbone:2013eqa,Carbone:2013rca,Carbone:2019hym}.}%
} 
By combining automated diagram generation and evaluation, we introduce here an automated approach for MBPT calculations of nuclear matter (at the two-body level).

\subsection{Automated approach for evaluating diagrams and sums of diagrams} \label{sec:acg_tda}

To evaluate the MBPT diagrams summarized in Table~\ref{tab:mbpt}, we optimize and extend the automated code generator (ACG) developed in Ref.~\cite{Drischler:2017wtt} to fifth and sixth order at the two-body level and third order at the three-body level, thereby implementing about 28154 additional diagrams.
The ACG translates each MBPT expression individually into optimized \CC source code, symbolically integrating out momentum-conserving delta distributions.
\rev{We emphasize again that the computational cost of evaluating MBPT at these high orders is not only determined by the rapidly increasing number of diagrams but also by the growing complexity of individual diagrams, including the dimensionality of the momentum integrals (see Table~\ref{tab:mbpt}) and their particle-hole and spin-isospin structure.}\footnote{%
\rev{See the companion GitHub repository~\cite{Drischler_Cheat_sheets} for the MBPT expressions and Appendix~\ref{app:comp_scaling} for the computational scaling of our MBPT implementation.}%
}

Complementary to the ACG approach, we have developed an improved approach based on a few code templates that can process tables of (multiple) MBPT expressions and optimizes their evaluation at runtime without generating explicit, diagram-specific source code.
In this context, the term ``table'' refers to a machine-readable data collection that encodes the compute graph structure of the MBPT diagram, including the particle-hole composition of the interaction vertices and energy denominator, and the overall factor of one or more MBPT expressions.
Our hybrid strategy uses the template-based approach for the grouped evaluation of MBPT diagrams beyond second order at the two-body level and applies the ACG approach to all other diagrams.

This template-driven approach is advantageous over the ACG, especially beyond third order, because it can evaluate sums of MBPT diagrams at a given MBPT order and level, whereas the ACG considers only one diagram at a time.
The new approach also enables more efficient GPU acceleration of the spin-isospin traces, which dominate the runtime at higher MBPT orders due to their exponential scaling with the number of particle-hole states.
In what follows, we discuss the advantages of the table-driven approach in more detail.

Evaluating sums of diagrams is motivated as follows.
First, beyond third order, MBPT expressions may appear in complex-conjugate pairs, and integrand functions may exhibit energy-denominator divergences that cancel order by order when added to the integrand function of the companion diagram. 
Reference~\cite{Wellenhofer:2021eis} discusses these divergences at fourth order in detail.
Here, we extended ADG to check for such divergences in diagrams at higher orders and make sure to group them accordingly.
Second, higher and more controllable numerical accuracies can be obtained due to partial cancellations among the terms in each group, whose uncertainties would otherwise be assumed uncorrelated and summed in quadrature.
These observed cancellations are consistent with previous findings~\cite{Holt:2016pjb} that suggest, at each MBPT order and each level separately, all diagrams should be evaluated consistently or not evaluated at all.
Third, computationally demanding tasks, such as normal ordering 3N forces across diagrams (see also Sec.~\ref{sec:chiral_forces}), can be combined and thus more efficiently parallelized, especially on GPUs.

We group MBPT diagrams order by order based on their particle-hole composition and then integrate the sums of their integrand functions.
The number of groups $N_\text{grp}  \ll N_\text{diag} $ for evaluating the $n$-th MBPT order in our calculations is summarized in Table~\ref{tab:mbpt}.
Beyond the Hartree-Fock level, we find that $N_\text{grp} = n$, except for $n=3$, where it is $N_\text{grp} = 2$.

Furthermore, we probe the sensitivity of our results to a regularization parameter that we add to the energy denominator of each integrand.
The sign of this parameter is chosen at each MBPT order such that the (removable) pole is moved outside the integration domain.
We efficiently perform MC integrations over a range of regularization parameters by rendering the integrand vector-valued, with each component associated with a different regularization parameter value.
Fitting for simplicity a quadratic polynomial to the components of this integration, then allows us to estimate the unregularized result.
We find that our results are robust to variations in the regularization parameter and extrapolation method.

The computational cost of evaluating the spin-isospin traces grows exponentially with the MBPT order.
At the two-body level, each additional order increases the number of spin-isospin states to be traced by a factor of 4, reaching, e.g.,  $2^{20}$ at fifth order and $2^{24}$ at sixth order.
Their computational cost at fifth order is already comparable to that for normal ordering 3N forces.  
Computationally efficient spin-isospin traces are therefore critical for our high-order MBPT calculations.

Our new template-driven approach utilizes runtime optimization techniques, including loop factorization and unrolling, compute-graph optimization, and GPU acceleration for spin-isospin traces. 
As a result of these optimizations, the spin-isospin traces become a negligible part of the integrand evaluation, even at high MBPT orders.
Specifically, we have derived, for each diagram's table entry, the subset of isospin configurations consistent with isospin conservation at each vertex. 
This yields a maximum count of 36 and 72 configurations across all diagrams at fifth and sixth order, respectively, representing a significant improvement.
Within the outer loop over valid isospin configurations, a 16-entry table is saved for each vertex with matrix elements corresponding to the 16 possible spin configurations of connected propagators, with the values precalculated before the inner spin configuration loop.  
The inner spin summation is then reorganized by finding the most connected vertex pair.   
The spins not involved with that pair are iterated over in an outer loop, which handles all non-pair vertex calculations outside the inner loop, improving efficiency, and the involved spins are iterated over in an inner unrolled loop of length 2 (found at fifth order and above), 4, or 8, depending on the number of connections between the vertices in the pair, producing the contribution of the pair. 

\subsection{Implementing chiral forces and normal ordering} \label{sec:chiral_forces}

We implement chiral interactions directly in a single-particle basis following the automated approach developed in Ref.~\cite{Drischler:2017wtt}. 
The $A$-body interactions are represented exactly as $4^A\times 4^A$ matrices in the spin-isospin space, and their matrix elements are implemented in \CC as momentum-dependent functions based on their operatorial definition.
As in Ref.~\cite{Drischler:2017wtt}, our current implementation assumes that the antisymmetrizer and the regulator functions commute for performance optimization.
That is, the regulator functions are invariant under permutations of particle labels, which includes the standard nonlocal regulator functions of the form
\begin{align}
    f_\text{NN}(p) &= \exp \left[-\frac{p^n}{\Lambda_\text{NN}^n} \right] \,,\\
    f_\text{3N}(p,q) &= \exp \left[- \frac{(p^2 + 3 q^2/4)^2}{\Lambda_\text{3N}^4} \right]\,,
\end{align}
where $p$ and $q$ are the magnitudes of the Jacobi momentum, $n >0$ is the exponent, and $\Lambda_\text{NN}$ and $\Lambda_\text{3N}$ are the NN and 3N momentum cutoffs, respectively.
If the operatorial definition is not available, e.g., for renormalization group-evolved interactions, the matrix elements are obtained by accumulating their partial-wave contributions up to a total momentum $J_\mathrm{max}=7$ at given momenta.
We have implemented this partial wave-based approach only at the NN level.
Hence, our implementation is compatible with, e.g., the Hebeler~\etal\ interactions~\cite{Hebeler:2010xb}, including the so-called magic interaction, and those developed in Ref.~\cite{Drischler:2017wtt} (here referred to as DHS) by adjusting the two 3N low-energy couplings at N$^2$LO to the triton binding energy and the empirical saturation point in SNM.

Baryon number conservation of the nuclear interactions makes these interaction matrices sparse, as only up to $4^{A} \; \binom{2A}{A}$ matrix elements can be non-zero~\cite{Drischler:2017zkp}.\footnote{The density of the interaction matrices corresponds to approximately 38\%~($A=2$), 31\%~($A=3$), and 27\%~($A=4$).}
Furthermore, due to the symmetries of the nuclear interactions, only a subset of the remaining matrix elements needs to be computed from which the entire matrix can be efficiently reconstructed; e.g., the chiral 3N interactions we consider do not break isospin symmetry. 
These physics-based optimizations speed up our many-body calculations by about an order of magnitude compared to the implementation in Ref.~\cite{Drischler:2017wtt}.

We have developed a hybrid CPU--GPU approach for normal ordering that allows us to evaluate the effective two-body interaction~\eqref{eq:Veff} on the fly for given external momenta.\footnote{Another approach would be to expand the vertices involving the effective two-body potential~\eqref{eq:Veff} and evaluate the NN and 3N terms separately at the expense of inflating $N_\text{diag}(n)$; e.g., by a factor of $2^n$ at $n$th order at the two-body level. See Refs.~\cite{Hebeler:2009iv,Drischler:2017wtt} for details.}
While the CPUs compute the NN contribution, the GPUs normal order the 3N forces in parallel CUDA streams\footnote{NVIDIA CUDA streams enable the numerous threads on the CPUs to time-share the few available GPUs efficiently.} using a combination of Gauss--Legendre and Lebedev~\cite{Lebedev:1999AQF} quadrature for the three-dimensional momentum integrals.
We do not apply approximations on the total momentum of the normal ordered 3N forces, i.e., $\vb{P}=0$~\cite{Hebeler:2009iv,Holt:2019bah,Hebeler:2020ocj} or averaged over all directions~\cite{Drischler:2015eba}, and instead rely on GPU acceleration across multiple devices to keep the calculations tractable.
By retaining the full $\vb{P}$ dependence, our framework will enable future studies to test these approximations quantitatively and to elucidate why they are typically good approximations.

Hence, the normal-ordered 3N contributions are numerically exact up to the (adjustable) accuracy of the adjustable momentum quadrature.
To optimize our calculations for both accuracy and runtime, we determine a set of suitable quadrature rules calibrated to the results of a highly accurate, adaptive cubature method~\cite{Johnson:2020Cubature}:
Specifically, for normal ordering, we choose here a 16-point Gauss--Legendre quadrature for the radial and an order-15 Lebedev quadrature rule for the angular integration to balance accuracy and computational speed. 
The order of the Lebedev rule refers to the maximum $\ell$ of the spherical harmonic $Y_\ell^m$ that can be exactly integrated by the rule.
Furthermore, we find that GPU acceleration can be optimized by normal-ordering all 3N interactions contributing to a diagram simultaneously, rather than individually, in a single asynchronous GPU transaction, thereby reducing GPU transaction overhead.
Asynchronous means that CPU portions of the computation continue while the GPU performs normal ordering for additional parallelization.
Overall, the GPU-accelerated normal ordering is about two orders of magnitude faster than a similar CPU-only calculation.

We obtain highly accurate results for the single-particle energy~\eqref{eq:spe} and HF energy~\eqref{eq:E_HF} by normal ordering the effective two-body interaction~\eqref{eq:Veff} repeatedly (typically $\lesssim 10 \keV$). 
Given a total nucleon density $n$ and isospin asymmetry $\delta $ (or proton fraction $x$) of nuclear matter, we sample the single-particle energy~\eqref{eq:spe} on a fine momentum grid only once in a preprocessing step and then use a cubic spline interpolator (for the self-energy) in our subsequent MBPT calculations.

\subsection{MC integration: \pvegas and \mpijm} \label{sec:PVegas}

Monte Carlo (MC) integration is the tool of choice for evaluating the multidimensional momentum integrals considered in this work with controllable uncertainties.
Their dimensions are ${N_\text{dim} \leqslant 27}$ up to fourth order and reach up to $N_\text{dim} = 30$ at fifth order, as shown in Table~\ref{tab:mbpt}. 
Many modern MC integrators exist in the literature, including CUBA~\cite{Hahn:2004fe,Hahn:2005pf}, VEGAS enhanced~\cite{Lepage:2020tgj,Lepage:1977sw}, and FOAM~\cite{Jadach:2003,Jadach:2007}.
Furthermore, Refs.~\cite{Brady:2021plj,Wen:2024shw} have recently demonstrated that normalizing flows enable efficient importance sampling estimators for MBPT calculations of the finite-temperature EOS. 
Here, we implement an improved version of Lepage's adaptive VEGAS algorithm~\cite{Lepage:1977sw}, named \pvegas, for adaptive variance reduction via importance sampling, optimized for the computationally challenging task at hand.  

\pvegas is a \CC library for high-performance MC integration of vector integrands, which has been extensively tested on a wide range of compute farms, including Summit at the Oak Ridge Leadership Computing Facility (OLCF).
Both random number generation (Sobol or Mersenne--Twister sequences) and integrand evaluation are distributed across multiple threads and compute nodes via OpenMP, MPI, and CUDA.
The latter facilitates the parallelized computation of one or more integrand functions on NVIDIA GPUs. 
Specifically, we use GPUs in our nuclear matter calculations to accelerate the computation of normal ordered 3N contributions, as discussed in Secs.~\ref{sec:acg_tda} and~\ref{sec:chiral_forces}.
We plan to make \pvegas publicly available as open-source software.

\pvegas divides the MC integration into a training and an evaluation phase.
During the training phase, the variance is reduced by iteratively adjusting the importance sampling distribution, similar to VEGAS, but with an adaptive number of sampling points per iteration.
Once a training accuracy goal is achieved, the importance-sampling distribution is frozen and saved externally for later use, enabling one to restart the integration or accelerate convergence at a neighboring point in the integrand's parameter space; e.g., variations in density and low-energy couplings (see also Ref.~\cite{Brady:2021plj}).
We find that the importance-sampling distributions for a given MBPT calculation propagate well in density, provided the variation is not too large (e.g., $\Delta n \lesssim 0.02 \fmiq$), and can then be quickly updated with an abbreviated training phase.
This feature allows us to map the density dependence of the nuclear EOS more efficiently.

Because the MBPT integrands are rotationally invariant, we can fix the direction (i.e., the polar and azimuthal angle) of one momentum vector and the azimuthal angle of another.
The corresponding integrals may then be carried out analytically, lowering their dimensionality $N_\text{dim}(n)$, as given in Table~\ref{tab:mbpt}, by three. 
As expected, this reduction somewhat tames the curse of dimensionality at high orders in the MBPT expansion, leading to a slightly accelerated numerical convergence.

We use the MPI-based job manager \mpijm~\cite{Berkowitz:2017xna,Berkowitz:2018gqe}, to efficiently orchestrate the parallel execution of the numerous \pvegas integrations to map the density and asymmetry dependence of the nuclear EOS at high MBPT orders efficiently. 
\mpijm monitors multi-node and multi-GPU compute jobs, automatically relaunching (or resuming) those that failed to complete before the wall time, while respecting job dependencies, including those arising from propagated importance sampling distributions. 
This approach is significantly more efficient on supercomputers than submitting smaller jobs individually.

Guided by parameter-based runtime estimates (e.g., based on MBPT order, proton-fraction, and error goal), \mpijm determines the number of nodes to control wall clock time for each job and creates a scheduling queue.  
Each job is, in turn, launched on a set of nodes clustered within the supercomputer network to enable fast MPI communication.  
The scheduling algorithm accounts for a job's dependency on the importance-sampling output of another job, e.g., by computing a neighboring density while holding other parameters constant.
Appendix~\ref{app:comp_scaling} provides more details on the computational scaling. 
The result of each completed compute task is recorded in an SQL database~\cite{Chang:2019khk} for data analysis.

Taken together, these advanced features enable large-scale MBPT calculations of infinite matter on leadership computing facilities.

\section{\rev{MBPT(5) results based on the Hebeler~\etal interactions}}
\label{sec:results}


We compute the energy per particle $E(n, x)/A$ as a function of the total nucleon density $n = \iso{n}{n} + \iso{n}{p}$ and proton fraction $x = \iso{n}{p}/n$, with the proton density $\iso{n}{p}$ and neutron density $\iso{n}{n}$, respectively.
The proton fraction is related to the isospin asymmetry (or neutron excess) via $\delta = 1-2x$.
\rev{All other quantities are derived from these calculations.} 

As discussed in Sec.~\ref{sec:acg_tda}, MBPT diagrams of a given order are evaluated in groups defined by their particle-hole content.
The MC integrator is configured with an accuracy goal of $\leqslant 20 \keV$, chosen as a compromise between accuracy and computational cost, and with a fixed initial number of MC samples per iteration.
Both settings are kept independent of density and interactions.
Because the convergence rate depends on the interaction and the diagram order, exploratory runs were performed to determine the initial sample size based on the slowest-converging group. 
Diagram groups that converge more rapidly typically achieve substantially smaller numerical errors and therefore contribute less to the total error.

An OLCF Director's Discretionary project award enabled us to run these computationally expensive calculations on the now-decommissioned supercomputer Summit.
Each Summit node consisted of two IBM POWER9 22-core CPUs and six NVIDIA Tesla V100 GPUs, which our MBPT framework fully leveraged, achieving nearly 100\% GPU utilization due to high GPU-driven parallelization.
A typical \mpijm batch run involved 128 nodes for about eight hours and hundreds of \pvegas jobs, while a typical job (i.e., one diagram group at a specific proton fraction and density and for one interaction) involved eight nodes with six MPI ranks per node, one GPU per rank, and 28 OpenMP threads per rank, for a total of 1344 threads.
In light of these findings, we expect that our MBPT framework will perform similarly well on current and next-generation supercomputers, such as Frontier at OLCF.

\subsection{\rev{PNM and SNM}}
\label{sec:pnm_snm_mbpt5}

\begin{table}[tb]
\renewcommand{\arraystretch}{1.2}
\setlength{\tabcolsep}{3pt}
    \centering
    \caption{%
    First set of interactions considered here:
    The nonlocal Hebeler~\etal\ interactions are comprised of SRG-evolved \NNNLO NN interactions and unevolved \NNLO 3N interactions.
    The SRG resolution scale is denoted by $\lambda$, the 3N momentum cutoff by $\Lambda_\textrm{3N}$, and the 3N LECs by $c_D$ and $c_E$. 
    The $\pi$N low-energy couplings $c_{i}$ match those in the NN potential (EM~$500\MeV$), except for the sixth potential, which uses the values obtained from the NN partial-wave analysis (PWA) in Ref.~\cite{Rentmeester:2003mf}. 
    Following the discussion in Section IV~B of Ref.~\cite{Drischler:2015eba}, we exclude one of the interactions developed in Ref.~\cite{Hebeler:2010xb}.
    We refer to these potentials using the label ``$(\lambda_\mathrm{SRG}/\Lambda_\mathrm{3N})$.''
    See also Refs.~\cite{Hebeler:2010xb,Drischler:2015eba} and the main text for details.%
}
    \label{tab:hebeler_interactions}
    \begin{ruledtabular}
    \begin{tabular}{d{1.1}d{1.1}Sld{1.3}d{1.3}}
         \multicolumn{1}{Sc}{$\lambda_\mathrm{SRG}$ [$\fmi$]} &
         \multicolumn{1}{Sc}{$\Lambda_\textrm{3N}$ [$\fmi$]} & 
         \multicolumn{1}{Sc}{3N $c_i$'s} & 
         \multicolumn{1}{Sc}{$c_D$} & 
         \multicolumn{1}{Sc}{$c_E$}  \\
        \hline
        1.8 & 2.0 & EM~$500\MeV$ & +1.264  & -0.120\\
        2.0 & 2.0 & EM~$500\MeV$ & +1.271  & -0.131\\
        2.0 & 2.5 & EM~$500\MeV$ & -0.292  & -0.592\\
        2.2 & 2.0 & EM~$500\MeV$ & +1.214  & -0.137\\
        2.8 & 2.0 & EM~$500\MeV$ & +1.278  & -0.078\\
        2.0 & 2.0 & PWA\footnote{We call this potential ``$(2.0/2.0)^*$'' to indicate the different choice for the 3N $c_i$'s.} & -3.007  & -0.686
    \end{tabular}
    \end{ruledtabular}
\end{table}

The first set \rev{of interactions} is comprised of the six commonly used Hebeler~\etal\ interactions, including the so-called ``magic'' interaction labeled ``(1.8/2.0).''
These soft interactions were obtained in Ref.~\cite{Hebeler:2010xb} by evolving the \NNNLO NN potential EM~$500\MeV$~\cite{Entem:2003ft} to lower resolution scales $\lambda_\mathrm{SRG}$ using the similarity renormalization group (SRG) and then combining them with unevolved \NNLO 3N forces.
The 3N LECs $c_D$ and $c_E$ were fit to the \isotope[3]{H} binding energy and the \isotope[4]{He} charge radius for two 3N momentum cutoffs $\Lambda_{\mathrm{3N}}$.
Table~\ref{tab:hebeler_interactions} summarizes the parameters of these low-momentum interactions.
We choose these interactions because they are SRG-evolved and with relatively low 3N cutoffs (i.e., softer), which improves the MBPT convergence significantly~\cite{Hebeler:2010xb,Huther:2019ont,Arthuis:2024mnl}, and because of their good reproduction of ground-state energies and charge radii in the medium-mass region~\cite{Simonis:2017dny,Hoppe:2019uyw}.
The ``(1.8/2.0)'' interaction in particular has shown an excellent reproduction of ground- and excited-states energies and has been widely used by practitioners~\cite{Stroberg2021a,Miyagi:2021pdc,Bonaiti2025}, but tends to slightly underpredict charge radii~\cite{Simonis:2017dny,Arthuis:2024mnl}.

\begin{figure}[tb]
    \centering
    \includegraphics[width=\columnwidth]{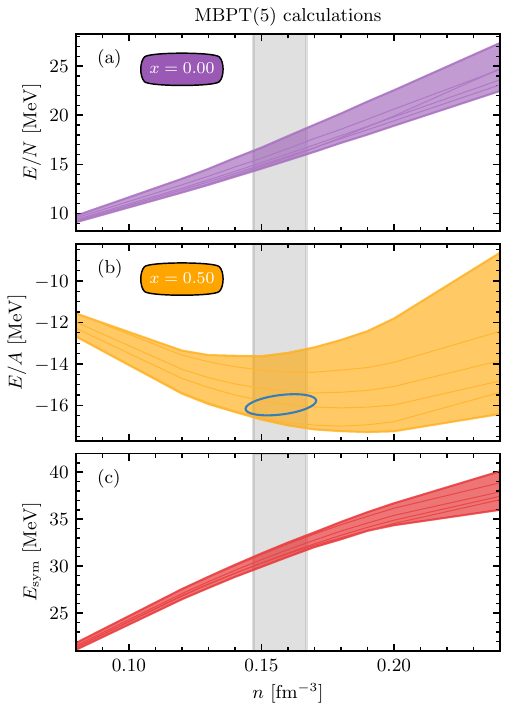}
    \caption{%
    Energy per particle in PNM~(a) and SNM~(b), and the symmetry energy~\eqref{eq:esym}~(c) based on MBPT(5) and the six Hebeler~\etal\ interactions in Table~\ref{tab:hebeler_interactions}.
    Each line corresponds to one of these Hamiltonians, and their spread at a given density, indicated by the bands, serves as a simple uncertainty estimate.
    The diagrammatic content of these calculations is summarized in Table~\ref{tab:mbpt}.
    All NN and 3N contributions up to fifth order at the normal-ordered two-body level and residual 3N contributions up to third order are included.
    The empirical saturation point, a bivariate student-$t$ distribution inferred in Ref.~\cite{Drischler:2024ebw} from a wide range of density functionals, is depicted by the blue ellipse, which encompasses the 95\% confidence region. 
    To guide the eye, the corresponding marginalized empirical saturation density at the 95\% confidence level is shown by the gray vertical bands.
    The extracted saturation points in SNM are depicted in Fig.~\ref{fig:coester_plot}. %
    } \label{fig:eos_pnm_snm_5th}
\end{figure}

Figures~\ref{fig:eos_pnm_snm_5th}(a) and~(b) depict the corresponding energy per particle in PNM and SNM, respectively, based on our MBPT calculations up to fifth order (``MBPT(5)'') at $n = 0.08 - 0.24 \fmiq$, or $n \approx (0.5 -1.5) n_0^\star$ in terms of the canonical value for the empirical saturation density, $n_0^\star = 0.16 \fmiq$.
We restrict our MBPT(5) calculations to this density range because of their high computational demands, as discussed in Appendix~\ref{app:comp_scaling}.
The diagrammatic content of these calculations is summarized in Table~\ref{tab:mbpt}.
All NN and 3N contributions are included up to fifth order at the normal-ordered two-body level, while residual 3N contributions are included up to third order.
The total numerical error from the MC integrations across all interactions is estimated to be $\lesssim 50 \keV$, despite the many contributing MBPT diagrams.
We are not aware of other MBPT calculations based on chiral NN and 3N forces at this high MBPT order.

In addition, Fig.~\ref{fig:eos_pnm_snm_5th}(c) shows the symmetry energy in the standard quadratic approximation,
\begin{equation} \label{eq:esym}
    E_\text{sym}(n) = \frac{E}{N}(n,x=0) - \frac{E}{A}(n,x=0.5) \,.
\end{equation}
We evaluate Eq.~\eqref{eq:esym} for each Hamiltonian independently, thereby accounting for correlations between the EOS in the limits of PNM and SNM.
As one can see, because of these correlations, the symmetry energy~\eqref{eq:esym} is more tightly constrained than the in-quadrature sums of the associated uncertainties in the energies per particle.
See also Refs.~\cite{Drischler:2020yad,Drischler:2020hwi} for detailed discussions.

The lines in each panel of Fig.~\ref{fig:eos_pnm_snm_5th} correspond to one of the underlying interactions. 
Following previous work (e.g., in Refs.~\cite{Drischler:2015eba,Somasundaram:2020chb}), we treat the spread of these six Hamiltonians at a given density, which is depicted by the shaded bands, as a simple uncertainty estimate (although with unspecified statistical coverage).\footnote{%
See Ref.~\cite{Plies:2025vaq} for a different approach based on the singular value decomposition to quantify theoretical uncertainties in SRG-evolved interactions.%
}
Figure~\ref{fig:eos_pnm_snm_5th} shows that this spread is mainly determined by two Hamiltonians: 
``(2.0/2.0)*'' sets the upper and ``(1.8/2.0)'' the lower boundary, consistent with Refs.~\cite{Drischler:2015eba,Drischler:2017wtt}.

The blue ellipse in Fig.~\ref{fig:eos_pnm_snm_5th}(b) represents the posterior distribution of the empirical saturation point estimated in Ref.~\cite{Drischler:2024ebw} using a Bayesian mixing model and a wide range of Skyrme and relativistic mean-field (RMF) energy functionals at the 95\% confidence level.
This distribution of the empirical saturation point is given by the bivariate student-$t$ distribution
\begin{subequations}  \label{eq:emp_sat_point}
    \begin{align}
        \pr\left(n_0^\mathrm{(emp)}, E_0^\mathrm{(emp)} \right) &= t_{\nu=9} (\vb*{\mu}, \vb*{\Psi}) \,, \\
        \vb*{\mu} &\approx \begin{pmatrix}
        0.157 \\ -15.968
        \end{pmatrix}\,,\\
        \vb*{\Psi} &\approx\begin{pmatrix}
        0.004569^2 & 0.017481^2 \\ 0.017481^2 & 0.174918^2
        \end{pmatrix} \,,
    \end{align}
\end{subequations}
corresponding to Equation~(31) in Ref.~\cite{Drischler:2024ebw}, although with additional decimal places given for more accuracy.
We extracted Eq.~\eqref{eq:emp_sat_point} from the open-source software accompanying Ref.~\cite{Drischler:2024ebw}. 
To guide the eye, the marginalized distribution characterizing the empirical saturation density, represented by the univariate student-$t$ distribution $t_{\nu=9} (\mu = \vb*{\mu}_1, \sigma^2=\vb*{\Psi}_{11}) $, is illustrated at the 95\% confidence level by the gray vertical bands in Fig.~\ref{fig:eos_pnm_snm_5th} and throughout the paper.
At the 95\% confidence level, we obtain the EOS constraints:
\begin{subequations}
    \begin{align}
    \frac{E}{N}\left(n_0^\mathrm{(emp)}, x=0.0\right) &\approx 14.6-18.2\MeV \, ,\\
    \frac{E}{A}\left(n_0^\mathrm{(emp)}, x=0.5\right) &\approx -(13.4-17.1) \MeV \,,\\
    E_\text{sym}\left(n_0^\mathrm{(emp)}\right) &\approx 30.0-33.1 \MeV \,.
\end{align}
\end{subequations}
%

\subsection{Nuclear saturation point}
\label{sec:nuclear_saturation}

Next, we examine the predicted nuclear saturation point $(n_0,E_0/A)$, i.e., the minimum of the EOS in the limit of SNM, along with its many-body convergence:
\begin{subequations}
\begin{align}
    \pdv{n} \frac{E}{A}(n,x) \bigg|_{\substack{n=n_0\\x=1/2}} &= 0 \,,\\ 
    \frac{E}{A}(n,x)\bigg|_{\substack{n=n_0\\x=1/2}} &\equiv \frac{E_0}{A} \,.
\end{align}
\end{subequations}
%

\begin{figure}[tb]
    \centering
    \includegraphics{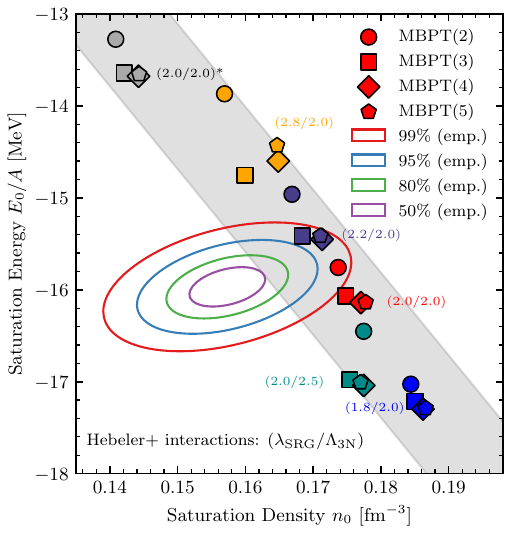}
    \caption{%
        Correlation between the predicted saturation density $n_0$ and saturation energy $E_0/A$ for the six Hebeler~\etal\ NN and 3N interactions (grouped by colors) obtained at second, third, fourth, and fifth order in the MBPT expansion (different markers). 
        The values for interaction's SRG resolution scale and 3N momentum cutoff are annotated as ``$(\lambda_\mathrm{SRG}/\Lambda_\mathrm{3N})$'' in units of $\fmi$ (see also Table~\ref{tab:hebeler_interactions}).
        These results correspond to the SNM calculations depicted in  Fig.~\ref{fig:eos_pnm_snm_5th}(c).
        The gray band depicts the Coester band obtained by a linear least-squares fit to all results shown. 
        Its width is determined by the minimum width that covers all results.
        The colored ellipses represent the empirical saturation point estimated in Refs.~\cite{Drischler:2024ebw} based on Bayesian model mixing across a wide range of energy-density functionals at four confidence levels (see the legend).
        The associated bivariate student-$t$ distribution is given by Eq.~\eqref{eq:emp_sat_point}.%
} \label{fig:coester_plot}
\end{figure}

Figure~\ref{fig:coester_plot} shows our predictions for $(n_0, E_0/A)$ at different orders in the MBPT expansion (grouped by colors), corresponding to the energies per particle shown in Fig.~\ref{fig:eos_pnm_snm_5th}(b).
The annotations state the values for the interaction's SRG resolution scale and 3N momentum cutoff ``$(\lambda_\mathrm{SRG}/\Lambda_\mathrm{3N})$'' in units of $\fmi$ (see also Table~\ref{tab:hebeler_interactions}).
The different markers denote the predictions at second, third, fourth, and fifth order in the MBPT expansion, respectively (see the legend).

We extract the $(n_0, E_0/A)$ in Fig.~\ref{fig:coester_plot} using Gaussian Processes (GPs)~\cite{rasmussen2006gaussian}.
Numerical uncertainties estimated by the MC integration of the MBPT diagram groups are summed in quadrature (i.e., they are assumed to be independent) and taken into account when optimizing the hyperparameters of the GP's kernel.
Here, we use the zero-mean function and a radial basis function (RBF) kernel with an additional white-noise kernel.
The variance of the white-noise kernel is optimized during GP training and typically converges to values on the order of $10^{-16}$ or lower, indicating a negligible noise contribution.
Both the zero-mean function and the RBF kernel are common choices in machine learning~\cite{rasmussen2006gaussian}.
Furthermore, we follow Ref.~\cite{Drischler:2020hwi} and enforce that the numerical uncertainty has a lower bound of $20\keV$ to be more conservative. 
For each Hamiltonian and nuclear interaction, we train a GP, from which we obtain straightforwardly the $(n_0, E_0/A)$ depicted in Fig.~\ref{fig:coester_plot}.

As shown in Fig.~\ref{fig:coester_plot}, the differences in $(n_0, E_0/A)$ between successive MBPT orders decrease monotonically, demonstrating excellent many-body convergence of the predicted saturation point. 
This convergence pattern is somewhat slower for the hardest interaction, ``$(2.8/2.0)$,'' reflecting its larger SRG resolution scale. 
For example, the MBPT($n$) predictions based on ``(2.8/2.0)'' are $n_0 \approx 0.157$ ($n=2$), $0.158$ ($n=3$), $0.165$ ($n=4$), $0.165 \fmiq$ ($n=5$) in contrast with $0.184$ ($n=2$), $0.185$ ($n=3$), $0.186$ ($n=4$), $0.186 \fmiq$ ($n=5$) for the softer ``(1.8/2.0).''
Nevertheless, these results indicate that these SRG-evolved interactions are perturbative near the saturation density.

The gray band in Fig.~\ref{fig:coester_plot} represents the anti-correlation of $(n_0, E_0/A)$ known as the Coester band~\cite{Coester:1970ai}.
Its center was obtained by a linear least-squares fit to all results shown in Fig.~\ref{fig:coester_plot}, and its width was determined by the minimal spread that covers all results.
This Coester band overlaps significantly with the empirical saturation point, whose distribution function is depicted by the ellipses associated with several confidence levels as specified in the legend. 
However, none of the interactions saturates inside the 95\% confidence region (blue ellipse) of the empirical constraint, not even at the highest orders considered here.
While the interaction ``$(2.0/2.0)$'' reproduces well the saturation energy and ``$(2.8/2.0)$'' the saturation density, neither interaction simultaneously reproduces the empirical saturation density and energy.
Given the MBPT convergence pattern observed in Fig.~\ref{fig:coester_plot}, we conclude that higher-order MBPT contributions cannot reconcile the statistically significant discrepancy between the predicted and empirical saturation points for these interactions.
See Refs.~\cite{Drischler:2021kxf,Drischler:2024ebw} for detailed discussions of this general trend of chiral NN and 3N interactions.

Assuming all MBPT(5) results for $(n_0,E_0/A)$ in Fig.~\ref{fig:coester_plot} are random draws from a common bivariate normal distribution, we obtain the following mean vector and covariance matrix:
\begin{equation} %
    \vb*{\mu} \equiv \begin{bmatrix}
    n_0\\
    E_0/A
    \end{bmatrix}
    \approx \begin{bmatrix}
    0.170 \\ -15.66
    \end{bmatrix} \,, \quad 
    \vb*{\Sigma} \approx \mqty[0.015^2 & -0.134^2 \\ -0.134^2 & 1.44^2] \,,
\end{equation}
where $n_0$ is in units of $\fmiq$ and $E_0/A$ in $\MeV$.
The associated Pearson correlation coefficient is $\rho \approx -0.93$, indicating a strong (Coester-like) anti-correlation~\cite{Drischler:2024ebw} of $n_0$ and $E_0/A$, as also evident from Fig.~\ref{fig:coester_plot}.

\subsection{Many-body convergence and residual 3N contributions up to third order}
\label{sec:mb_conv_res_heb}

\begin{table*}[tb]
\renewcommand{\arraystretch}{1.1}
\caption{
A sample of the order-by-order MBPT convergence for the two Hebeler~\etal\ interactions ``(2.0/2.0)'' and ``(2.8/2.0).'' 
See also Table~\ref{tab:hebeler_interactions} for details on the interactions.
The MBPT contributions are given in MeV, in both PNM and SNM, and at $n = 0.08$, $0.16$, and $0.30 \fmiq$.
The column labeled ``PD'' encodes the proton fraction and density using a short-hand identifier: 
``$\textrm{P}$'' (``$\textrm{S}$'') stands for PNM (SNM), and the two numbers correspond to the integer part of $n \times 100$.
For example, ``$\textrm{P}16$''  denotes a PNM calculation at density $n = 0.16 \fmiq$.
The remaining columns show the contributions at the indicated MBPT order, with the HF energy including the kinetic energy. 
At the second (``2nd'') and third orders (``3rd''), residual three-body contributions are included.  
The column ``Total'' gives the sum of all contributions up to fifth order (``5th'').  
Estimates of the statistical uncertainties from the MC integration are reported in parentheses. 
The three columns on the right provide different Pad{\'e} approximants $[n/m]$ that re-sum the MBPT expansion.%
}
\label{tab:eos}
\begin{tabular*}{\textwidth}{l}
\hline \hline \vspace{-4pt} \\
Hebeler~\etal\ (2.0/2.0) \\
\end{tabular*}
\begin{ruledtabular}
{ 
\begin{tabular}{lccccccccc} {PD} & {HF} & {2nd}& {3rd} & {4th} & {5th} & {Total} & {Pad{\'e}[2/2]} & {Pad{\'e}[3/2]} & Pad{\'e}{[2/3]} \\
\hline \\[-1.5ex]
P08 & \phantom{0}\phantom{-}9.88(00) & -0.70(01) & \phantom{-}0.11(01) & -0.05(01) & \phantom{-}0.01(01) & \phantom{0}\phantom{0}9.26(01) & \phantom{0}\phantom{-}9.26 & \phantom{0}\phantom{-}9.26 & \phantom{0}\phantom{-}9.26 \\
P16 & \phantom{-}16.57(00) & -0.80(01) & \phantom{-}0.10(01) & -0.06(01) & \phantom{-}0.01(01) & \phantom{0}15.82(02) & \phantom{-}15.81 & \phantom{-}15.81 & \phantom{-}15.81 \\
P30 & \phantom{-}29.39(00) & -0.85(01) & \phantom{-}0.06(01) & -0.09(01) & \phantom{-}0.03(02) & \phantom{0}28.53(02) & \phantom{-}28.51 & \phantom{-}28.52 & \phantom{-}28.52 \\
\hline
S08 & \phantom{0}-8.09(00) & -3.75(01) & -0.36(01) & -0.01(01) & -0.07(01) & -12.28(02) & -12.20 & -12.23 & -12.29 \\
S16 & -12.02(00) & -3.61(01) & -0.30(01) & -0.05(01) & \phantom{-}0.02(01) & -15.96(01) & -15.98 & -15.97 & -15.95 \\
S30 & \phantom{0}-7.24(00) & -3.50(01) & -0.31(01) & -0.31(01) & -0.03(03) & -11.39(03) & -11.45 & -11.42 & -11.42 \\
\end{tabular}
}
\end{ruledtabular}

\begin{tabular*}{\textwidth}{l}
\vspace{-3pt} \\
Hebeler~\etal\ (2.8/2.0) \\
\end{tabular*}

\vspace{2pt} \begin{ruledtabular}
\begin{tabular}{lccccccccc} {PD} & {HF} & {2nd}& {3rd} & {4th} & {5th} & {Total} & {Pad{\'e}[2/2]} & {Pad{\'e}[3/2]} & {Pad{\'e}[2/3]} \\
\hline \\[-1.5ex]
P08 & \phantom{-}10.77(00) & -1.43(01) & \phantom{-}0.28(01) & -0.11(01) & \phantom{-}0.03(01) & \phantom{0}\phantom{0}9.54(01) & \phantom{0}\phantom{-}9.53 & \phantom{0}\phantom{-}9.53 & \phantom{0}\phantom{-}9.53 \\
P16 & \phantom{-}18.68(00) & -2.30(01) & \phantom{-}0.38(01) & -0.18(01) & \phantom{-}0.01(02) & \phantom{0}16.60(02) & \phantom{-}16.62 & \phantom{-}16.58 & \phantom{-}16.59 \\
P30 & \phantom{-}34.32(00) & -3.62(01) & \phantom{-}0.32(01) & -0.40(02) & \phantom{-}0.05(02) & \phantom{0}30.68(03) & \phantom{-}30.65 & \phantom{-}30.63 & \phantom{-}30.63 \\
\hline
S08 & \phantom{0}-4.55(00) & -6.55(01) & -0.71(01) & \phantom{-}0.25(01) & -0.00(01) & -11.56(02) & -11.51 & -11.57 & -11.58 \\
S16 & \phantom{0}-5.75(00) & -8.07(01) & -0.90(01) & \phantom{-}0.16(01) & \phantom{-}0.15(02) & -14.40(02) & -14.51 & -14.31 & -14.38 \\
S30 & \phantom{0}\phantom{-}2.88(00) & -9.77(01) & -1.00(01) & -0.66(02) & \phantom{-}0.10(03) & \phantom{0}-8.46(04) & \phantom{0}-8.75 & \phantom{0}-8.53 & \phantom{0}-8.44 \\
\end{tabular}
\end{ruledtabular}
\end{table*}

Table~\ref{tab:eos} summarizes the order-by-order convergence pattern of our MBPT calculations up to fifth order for two representative Hebeler~\etal\ interactions, ``(2.0/2.0)'' and ``(2.8/2.0)'' at $n = 0.08$, $0.16$, and $0.30\fmiq$.
The column labeled ``PD'' encodes the proton fraction and density as follows: 
``$\textrm{P}$'' (``$\textrm{S}$'') stands for PNM (SNM), and the following number $D$ corresponds to the density $n = D/100 \fmiq$.
For example, ``$\textrm{P}16$''  denotes a PNM calculation at $n = 0.16 \fmiq$.
The next five columns show the contributions to the energy per particle at the HF level and from the second (``2nd''), third (``3rd''), fourth (``4th''), and fifth (``5th'') MBPT orders.
The HF results include the kinetic energies, and the second- and third-order results include the residual three-body contributions.
The column ``Total'' gives the total energy per particle at fifth order; i.e., the sum of all MBPT contributions up to this order.  
Estimates of the statistical uncertainties arising from MC integration are reported in parentheses.
To assess the convergence pattern further, we give in the last three columns the results of three Pad{\'e} approximants,\footnote{%
A Pad{\'e} approximant~\cite{baker1961pade} of a function $f(x)$ is a rational function, $[m/n](x) = P_m(x)/Q_n(x)$, where $P_m(x)$ and $Q_n(x)$ are polynomials of degrees $m$ and $n$, respectively.
It is constructed so that its Taylor series expansion about a given $x$ matches that of $f(x)$ up to order $k = m+n$.
For example, using MBPT(5) calculations ($k = 5$), the highest symmetric Pad{\'e} approximant one can construct is $[2/2]$.%
} 
Pad{\'e}$[2/2]$, $[3/2]$, and $[2/3]$, which provide simple rational-function resummations\footnote{%
See also, e.g., Refs.~\cite{bender1999advanced,Goodson:2011,Tichai:2020dna} for overviews of other resummation techniques, including eigenvector continuation~\cite{Frame:2017fah,Demol:2019yjt}.%
} 
of the MBPT expansion.
All energies are reported in MeV. 

In PNM, we observe a rapid order-by-order convergence, with the third- and fourth-order contributions typically yielding only a few $100 \keV$, while the fifth-order contributions are suppressed to only a few $10 \keV$. 
These higher-order contributions are small compared to the total energy per particle, numerical accuracy, and expected EFT truncation error of a few MeV at the highest density (see Sec.~\ref{sec:fit_interactions}). 
Recall that the third-order contributions include the 3N residual terms.
Across all densities, the three Pad{\'e} approximants agree well with the MBPT(5) results and show excellent consistency among themselves.

In SNM, as one might expect, the observed MBPT convergence is slower, especially for the harder (2.8/2.0) interaction at $n \geqslant 0.16 \fmiq$.
For example, at the highest $n =0.30 \fmiq$, we find that the
third- and fourth-order contributions are $\approx 0.5-1.0 \MeV$ in magnitude, and the fifth-order contributions are attractive and at the $100 \keV$ level for this interaction.
Nonetheless, our MBPT(5) results still agree well with the Pad{\'e} approximants within that energy range.

Overall, Table~\ref{tab:eos} shows that the MBPT contributions generally decrease monotonically with increasing order, indicating that these SRG-evolved interactions are perturbative. 
The observed convergence pattern suggests that MBPT up to fifth order provides a controlled expansion of the nuclear EOS in the considered density regime, consistent with previous findings~\cite{Drischler:2015eba,Drischler:2017wtt}.
Extending the calculations to MBPT(6) would provide further insights into the asymptotic behavior of MBPT at higher orders, particularly for stronger interactions and at high densities, and would enable the construction of higher Pad{\'e} approximants, including $[3/3]$. 
However, as we have observed explicitly, such calculations are computationally challenging, requiring roughly an order of magnitude more computational resources than the present fifth-order results (see also Appendix~\ref{app:comp_scaling}).

\begin{figure}[tb]
    \begin{center}
    \includegraphics[width=\linewidth]{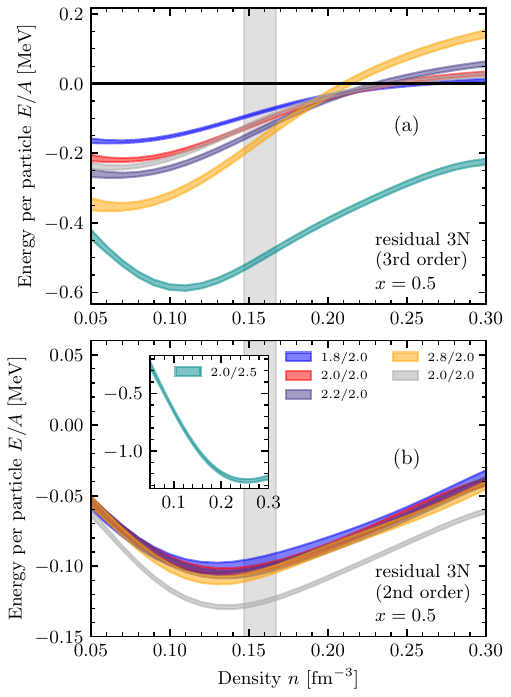}
    \end{center}
    \caption{%
    Residual 3N contributions to the energy per particle as a function of the density at third (top panel) and second order (bottom panel) in SNM based on the six Hebeler~\etal\ potentials (see Table~\ref{tab:hebeler_interactions}).
    The uncertainty bands represent the GPs discussed in the main text and correspond to the $1\sigma$ credibility region.
    The vertical bands depict the empirical range for the saturation density, corresponding to Eq.~\eqref{eq:emp_sat_point}.%
    }
    \label{fig:residual3N_3rd_snm}
\end{figure}

Figure~\ref{fig:residual3N_3rd_snm} shows the residual 3N contributions at third [Fig.~\ref{fig:residual3N_3rd_snm}(a)] and second order [Fig.~\ref{fig:residual3N_3rd_snm}(b)] in SNM for $n = 0.05 - 0.30 \fmiq$.
The bands depict the numerical uncertainties at the $1\sigma$ confidence level, and the gray vertical band corresponds to the empirical range for the saturation density.
We have verified that our second-order results, which are always attractive, agree with those reported in Ref.~\cite{Drischler:2017wtt}.
The third-order residual contributions in SNM are attractive at saturation density and in the range of $|E_\mathrm{res}^{(3)}(n_0)/A|\approx 0.1-0.5 \MeV$, where the upper limit is due to the interaction with the larger cutoff, $\Lambda_\mathrm{3N} = 2.5 \fmi$.
At $n \approx 0.22 \fmiq$, we find a zero crossing for the interactions with $\Lambda_\mathrm{3N} = 2.0 \fmi$.
In comparison, at second order, we find $|E_\mathrm{res}^{(2)}(n_0)/A|\lesssim 0.1 \MeV$ disregarding the interaction $\Lambda_\mathrm{3N} = 2.5 \fmi$, which, similar to our findings at third order, can have (significantly) larger attractive contributions; 
especially beyond the saturation density, as shown in the inset in Fig.~\ref{fig:residual3N_3rd_snm}(a).

Overall, we find that residual 3N contributions at both second and third order exhibit a significant cutoff dependence, which is stronger at second order for the interactions considered here. 
They are not generally suppressed at third order relative to second order, except in cases where the density happens to lie near the zero crossing shown in Fig.~\ref{fig:residual3N_3rd_snm}(a).
However, we find that the residual 3N contributions are significantly smaller than the second-order contributions at the normal-ordered two-body level (see Table~\ref{tab:eos}) and are likely also smaller than the EFT truncation error. 
This conclusion depends on the details of the underlying NN and 3N interactions and should therefore be validated for each specific application.

The corresponding residual contributions in PNM are smaller than those in SNM, with $|E/A| \ll 40 \keV$ for most interactions across all densities. 
They are presented in Appendix~\ref{app:add_results} for completeness.
The largest contributions in magnitude arise for the interaction with $\Lambda_\text{3N} = 2.5 \fmi$, further highlighting the cutoff dependence noted earlier. 
Hence, these contributions are negligible compared with the 3N contributions at the normal-ordered two-body level.

\section{\rev{Meta-modeling of neutron-rich matter using MBPT(5)}}
\label{sec:eos_modeling_framework}

Motivated by the disagreement between the predicted and empirical saturation point shown in Fig.~\ref{fig:coester_plot}, we follow here the parametric approach introduced in Ref.~\cite{Hebeler:2013nza} to study the EOS of neutron-rich matter and extract low-density EOS parameters from our MBPT(5) calculations shown in Fig.~\ref{fig:eos_pnm_snm_5th}.
This analytic EOS model allows us to incorporate empirical nuclear saturation properties into our analysis. 

\subsection{\rev{Bayesian parameter estimation}}

Specifically, we model the dependence of the energy per particle on $(n, x)$ using the meta-model~\cite{Hebeler:2013nza}
\begin{equation} \label{eq:eos_model}
\begin{split}
        \frac{\bar{\varepsilon}(n,x; \vb*{\mathcal{P}} )}{T_0} &= \frac{3}{5} \left[x^{5/3}+(1-x)^{5/3}\right] \bar{n}^{2/3} \\
&\quad- \left[(2 \alpha - 4 \alpha_L) x (1 - x) + \alpha_L \right]\bar{n} \\ 
&\quad + \left[(2 \eta - 4 \eta_L) x (1 - x) + \eta_L \right] \bar{n}^\gamma \,,
\end{split}
\end{equation}
with the SNM kinetic energy $T_0 = (3\pi^2 n_0^\star/2)^{2/3}/(2m) \approx 36.84 \MeV$ at $n_0^\star = 0.16 \fmiq$, averaged nucleon mass $m \approx 938.9 \MeV$, and scaled density $\bar{n} = n / n_0^\star$.
We fix the exponent $\gamma = 4/3$ following Ref.~\cite{Hebeler:2013nza}, which found that the predicted symmetry energy is only weakly sensitive to variations in the range $1.2 \leqslant \gamma \leqslant 1.45$ (see Table~1 in Ref.~\cite{Hebeler:2013nza}).
Once the model parameters $\vb*{\mathcal{P}}  = \{\alpha, \alpha_L, \eta, \eta_L; \gamma\}$ are constrained by our PNM calculations and the empirical saturation point, derived quantities can be calculated symbolically; e.g., using the Python library \texttt{SymPy}~\cite{sympy}. 
We consider here the following derived quantities:
The nuclear symmetry energy evaluated at $n_0$\footnote{Note that $S_v \neq E_\mathrm{sym}(n_0)$ in general because of the nonquadraticities induced by the kinetic term in the model~\eqref{eq:eos_model}.}
\begin{align}
S_v (\vb*{\mathcal{P}}) &= \frac{1}{8} \frac{\partial^2}{\partial x^2} \bar{\varepsilon}(n,x; \vb*{\mathcal{P}}) \bigg|_{\substack{n=n_0\\x=1/2}}\,, \label{eq:Sv}\\ 
\intertext{the associated slope parameter}
L (\vb*{\mathcal{P}}) &= \frac{3n_0}{8} \frac{\partial^3}{\partial n \partial x^2} \bar{\varepsilon}(n,x; \vb*{\mathcal{P}}) \bigg|_{\substack{n=n_0\\x=1/2}}\,, \label{eq:L}
\intertext{the incompressibility in SNM}
    K (\vb*{\mathcal{P}}) &= 9 n_0^2 \frac{\partial^2}{\partial n^2} \bar{\varepsilon}(n,x; \vb*{\mathcal{P}}) \bigg|_{\substack{n=n_0\\x=1/2}}\,, \label{eq:K}
\intertext{the pressure}
    p(n, x; \vb*{\mathcal{P}}) &= n^2 \frac{\partial}{\partial n} \bar{\varepsilon}(n,x; \vb*{\mathcal{P}}) \,,\label{eq:pressure}
\intertext{and the speed of sound squared}
	c_s^2(n,x; \vb*{\mathcal{P}}) &= 
    \dfrac{\frac{\partial p(n,x)}{\partial n}}{\left[ \left( 1 + n \frac{\partial}{\partial n} \right)
	\frac{E}{A}(n,x; \vb*{\mathcal{P}}) + m \right]} \,. \label{eq:cs2}
\end{align}
%

\begin{figure*}
    \centering
    \includegraphics[width=1\linewidth]{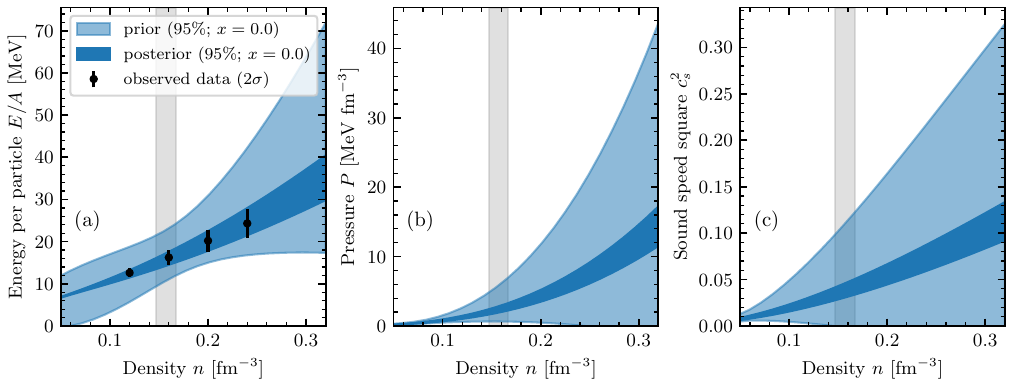}
    \caption{%
    Posterior (dark blue bands) and prior distributions (light blue bands) of the PNM EOS as predicted by the EOS model~\eqref{eq:eos_model}.
    Panel~(a) shows the energy per particle, panel~(b) the pressure, and panel~(c) the sound speed squared at the 95\% confidence level (see the legend).
    The corresponding nuclear matter parameters and model parameters are depicted in Fig.~\ref{fig:corner}.
    The $2\sigma$-error bars in panel~(a) show the four data points used to calibrate the EOS model~\eqref{eq:eos_model}.
    Note that the EOS prior is chiral EFT-agnostic and weakly informed, e.g., as is evident at low densities in panel~(a).
    The gray vertical bands represent the 95\% confidence interval for the empirical saturation density.%
    }
    \label{fig:eos_posterior}
\end{figure*}

We constrain the model parameters $\vb*{\mathcal{P}} $ using a Bayesian regression analysis of our MBPT(5) results in PNM (see Fig.~\ref{fig:eos}(a)), guided by physics-informed prior assumptions as detailed below.
According to Bayes' theorem, the posterior (probability) distribution function of $\vb*{\mathcal{P}}$ given the observed data $\vb*{\mathcal{Y}}$ (and additional information $\mathcal{I}$) is determined by the product of the likelihood function $\pr(\vb*{\mathcal{Y}} \given \vb*{\mathcal{P}}, \mathcal{I})$ and the prior distribution $\pr(\vb*{\mathcal{P}} \given \mathcal{I})$ divided by the Bayesian evidence $\pr(\vb*{\mathcal{Y}})$:
\begin{equation} \label{eq:bayes_theorem}
    \pr(\vb*{\mathcal{P}} \given \vb*{\mathcal{Y}}, \mathcal{I}) 
    = 
    \frac{\pr(\vb*{\mathcal{Y}} \given \vb*{\mathcal{P}}, \mathcal{I}) \,
    \pr(\vb*{\mathcal{P}} \given \mathcal{I})}{\pr(\vb*{\mathcal{Y}})} \,.
\end{equation}
Here, $\vb*{\mathcal{Y}}$ is the vector of observed PNM energies per particle at the $d=4$ equidistant grid points $n_i \in \{ 0.12, 0.16, 0.20, 0.24\} \fmiq$.
The associated uncertainty estimates and all model assumptions, such as the parametric model~\eqref{eq:eos_model}, are collected in $\mathcal{I}$.

For the likelihood function, we use the usual normal distribution, corresponding to the log-likelihood function
\begin{equation}
\begin{split}
    \ln \pr( \vb*{\mathcal{Y}} \given \vb*{\mathcal{P}}, \mathcal{I}) 
        &= -\frac{1}{2}\bigg[
        (\vb{x}-\vb*{\mu})^\top \vb*{\Sigma}^{-1} (\vb{x}-\vb*{\mu}) \\
        & \quad + \ln \det(\vb*{\Sigma})
        + d \ln(2\pi)
        \bigg].
\end{split} 
\end{equation}
Here, we assume that $\vb*{\mathcal{Y}}$ has a normal distribution; i.e., $\vb*{\mathcal{Y}} \sim \normal(\vb*{\mu}, \vb*{\Sigma})$, with $\vb*{\mu}$ and $\vb*{\Sigma}$ are respectively estimated by the sample mean and sample covariance matrix of the computed PNM energies per particle across the six Hamiltonians.
This procedure accounts for correlations in the modelled EOS with respect to density.

To construct the prior $\pr(\vb*{\mathcal{P}} \given \mathcal{I})$, we use our empirical knowledge of the saturation point and nuclear symmetry energy, encoded in the joint prior distribution:
\begin{multline} \label{eq:prior}
    \pr(n_0^\mathrm{(emp)}, E_0^\mathrm{(emp)}, S_v^\mathrm{(emp)}, L^\mathrm{(emp)}) \\
    = \pr(n_0^\mathrm{(emp)}, E_0^\mathrm{(emp)}) \times 
    \pr(S_v^\mathrm{(emp)}) \times \pr(L^\mathrm{(emp)}) \,.
\end{multline}
That is, we assume 
\begin{subequations}
    \begin{align}
    n_0^\mathrm{(emp)}, E_0^\mathrm{(emp)}/A &\sim t_\nu (\vb*{\mu}, \vb*{\Psi}) \,, \\
    S_v^\mathrm{(emp)} &\sim \mathcal{N}(\mu = 32, \sigma^2 = 3^2) \,,\\
    L^\mathrm{(emp)} &\sim \mathcal{N}(\mu = 60, \sigma^2 = 25^2)\,,
\end{align}
\end{subequations}
where the mean values $\mu$ and standard deviations $\sigma$ are given in units of $\MeV$, and the bivariate $t$-distribution is defined by Eq.~\eqref{eq:emp_sat_point}. 
The weakly-informed prior distributions for $(S_v^\mathrm{(emp)}, L^\mathrm{(emp)})$ are chosen such that they accommodate, at the $2\sigma$ level, the large mean values extracted from PREX--II~\cite{Reed:2021nqk}, and the typically smaller values predicted by microscopic EOS calculations. 

We obtain random draws of the prior $\pr(\vb*{\mathcal{P}} \given \mathcal{I})$ by sampling from the distribution~\eqref{eq:prior} and determine, for each sample, first $(\alpha, \eta)$ such that
\begin{subequations} \label{eq:prior_conditions}
\begin{align}
    \bar{\varepsilon}\left(n_0^\mathrm{(emp)}, x=0.5 ;\vb*{\mathcal{P}}\right) &= E_0^\mathrm{(emp)} / A\,, \\
    p\left(n_0^\mathrm{(emp)}, x=0.5 ; \vb*{\mathcal{P}}\right) &= 0 \,, 
\intertext{and then $(\alpha_L, \eta_L)$ such that}
    S_v \left(n_0^\mathrm{(emp)} ;\vb*{\mathcal{P}}\right) &= S_v^\mathrm{(emp)} \,,\\
    L\left(n_0^\mathrm{(emp)} ;\vb*{\mathcal{P}}\right) &= L^\mathrm{(emp)} \,,
\end{align}
\end{subequations}
as defined by Eqs.~\eqref{eq:eos_model},~\eqref{eq:Sv},~\eqref{eq:L}, and~\eqref{eq:pressure}.
Note that $(\alpha_L,\eta_L)$ do not contribute to SNM, so $(\alpha,\eta)$ are uniquely determined by the empirical saturation point (in SNM).
The Markov chain Monte Carlo (MCMC) sampling is performed using the Python library \texttt{PyMC}~\cite{abrill-pla2023pymc}, resulting in the prior distribution $\pr(\vb*{\mathcal{P}} \given \mathcal{I})$ depicted in Appendix~\ref{app:add_results} for the interested reader.
We find that $\alpha$ and $\eta$, as well as $\alpha_L$ and $\eta_L$, are strongly correlated with one another.
Our prior distribution is compatible with $\alpha=5.87$ and $\eta=3.81$, obtained in Ref.~\cite{Hebeler:2013nza} by fitting them to the canonical point estimate $n_0^\star = 0.16 \fmiq$ and $ E_0^\star/A=-16 \MeV$. 
By construction, our prior distribution, however, explores a significantly larger region of the parameter space.
With our choice for $\gamma$, these prior assumptions constrain $K \approx 235 \pm 2 \MeV$ at the $1\sigma$ level, in agreement with the empirical value $K^\mathrm{(emp)} \approx 240 \pm 20 \MeV$ from Ref.~\cite{Roca-Maza:2018ujj}.
Note that these priors are chiral EFT-agnostic, as our microscopic EOS calculations have not informed them.

\subsection{Posterior distributions}
\label{sec:eos_modeling_posteriors}

With the statistical framework set up, we are now in the position to implement and sample from the posterior~\eqref{eq:bayes_theorem} using \texttt{PyMC}~\cite{abrill-pla2023pymc}.
Our results for the model parameters in Eq.~\eqref{eq:eos_model} are depicted in Fig.~\ref{fig:meta_model_posterior_params} in Appendix~\ref{app:add_results}.
We find that our posterior~\eqref{eq:bayes_theorem} explores a statistically consistent, but significantly larger, parameter space than that obtained in Ref.~\cite[cf. Figure~3]{Hebeler:2013nza}.
For each of the samples drawn, we evaluate the energy per particle~\eqref{eq:eos_model}, pressure~\eqref{eq:pressure}, and sound speed squared~\eqref{eq:cs2} in PNM.
Figure~\ref{fig:eos_posterior} shows the resulting posterior (dark blue bands) and associated prior (light blue bands) for $n = 0.05 - 0.32 \fmiq$ at the 95\% credibility level.
Note that this density region extends beyond the support of the data used for model calibration (depicted by the error bars), both to higher and lower densities. 
As can be seen in Fig.~\ref{fig:eos_posterior}, the EOS prior considered here (light blue bands) is only weakly informed, supporting a wide range of low-density EOSs at the 95\% confidence level, even allowing for somewhat bound ($E/N < 0$ at $n \lesssim 0.1 \fmiq$) or thermodynamically unstable EOSs ($P < 0$ or $c_s^2 < 0$ at $n \gtrsim 0.1\fmiq$) at this high credibility level.
Calibrated to our MBPT calculations, the posterior significantly narrows the uncertainty band, yielding both unbound and thermodynamically stable EOSs as expected.
For example, at 95\% confidence level, the prior supports $E/N \approx (17.3-71.9) \MeV$ at the highest density shown, $n = 0.32 \fmiq$, whereas the posterior significantly shrinks this band to $E/N \approx (29.9-40.4) \MeV$.
At the 95\% confidence level and the empirical saturation density, the EOS posterior predicts these ranges in PNM:
\begin{subequations}
    \begin{align}
        \frac{E}{N} \left(n_0^\mathrm{(emp)}\right) &\approx 14.06^{+1.87}_{-1.72} \MeV \,,\\ 
        p\left(n_0^\mathrm{(emp)}\right) &\approx 1.83^{+0.64}_{-0.55} \MeV \fmiq \,,\\
        c_s^2\left(n_0^\mathrm{(emp)}\right) &\approx 0.03 \pm 0.01 \,.
    \end{align}
\end{subequations}
%

\begin{figure*}[tbp]
    \centering
    \includegraphics[width=\textwidth]{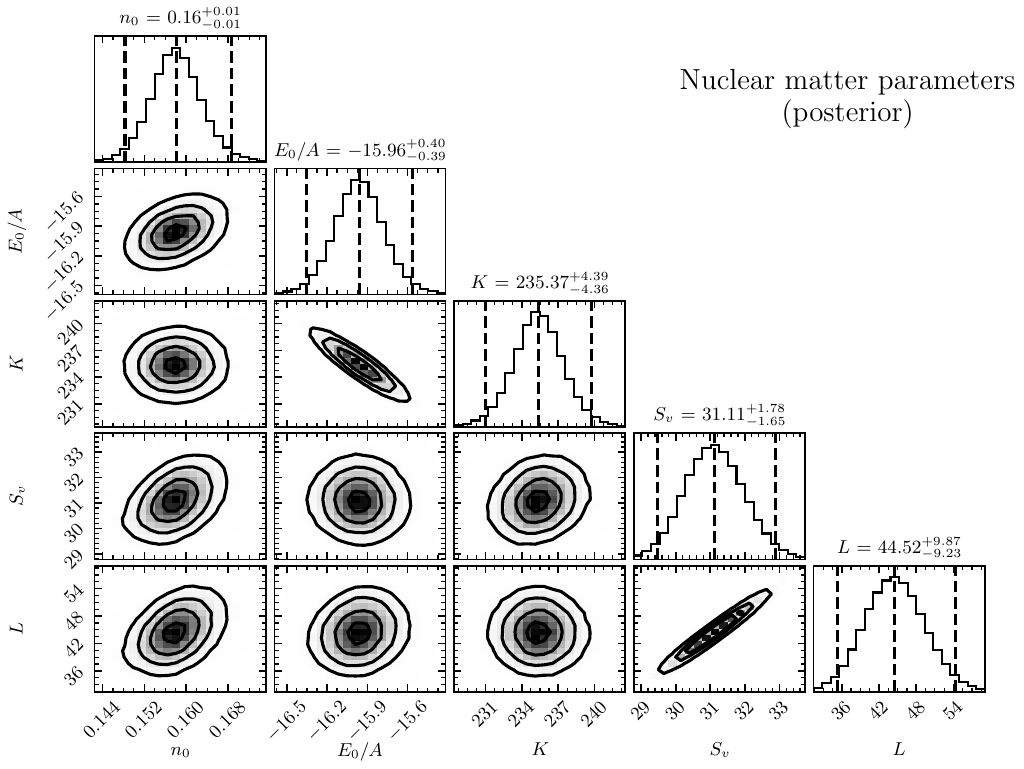}
    \caption{%
        Posterior distribution of the inferred nuclear matter parameters, obtained from our Bayesian inference based on the EOS model~\eqref{eq:eos_model}.
        The parameters are the nuclear saturation point $(n_0, E_0/A)$, the incompressibility $K$, and the symmetry energy $S_v$ and its slope parameter $L$ evaluated at $n_0$. 
        All parameters are in $\MeV$, except for $n_0$, which is in $\fmiq$.
        The titles on the diagonal panels state the 95\% confidence regions centered on the medians of the corresponding marginal distributions (dashed vertical lines), all of which fall within canonical ranges. 
        By construction, the joint distribution of $(n_0,E_0/A)$ approximates the empirical saturation point~\eqref{eq:emp_sat_point} to which it was calibrated. 
        The contour lines, moving outward, enclose the $0.5\sigma$, $1\sigma$, $1.5\sigma$, and $2\sigma$ confidence regions. %
    }
    \label{fig:corner}
\end{figure*}

Figure~\ref{fig:corner} shows the posterior distribution of the predicted nuclear matter parameters $n_0$, $E_0$, $K$, $S_v$, and $L$ at the posterior level.
The titles of the diagonal panels indicate the 95\% confidence regions for the corresponding marginal distributions (dashed vertical lines).
All parameters are given in $\MeV$, except for $n_0$, which is in $\fmiq$.
As can be seen in Fig.~\ref{fig:corner}, the model~\eqref{eq:eos_model} predicts $(S_v, L)$ to be strongly correlated and $(E_0, K)$ to be strongly anti-correlated.
Furthermore, the joint distribution of $(n_0,E_0/A)$ approximates the empirical saturation point~\eqref{eq:emp_sat_point} to which it was calibrated.
At the 95\% credibility level, we find for the symmetry energy in the normal approximation (i.e., $S_v, L \sim \normal (\vb*{\mu}, \vb*{\Sigma})$)
\begin{equation} %
    \vb*{\mu} \equiv \begin{bmatrix}
    S_v\\
    L
    \end{bmatrix}
    \approx \begin{bmatrix}
    31.1 \\ 44.6
    \end{bmatrix} \,,  \quad 
    \vb*{\Sigma} \approx \mqty[0.9^2 & 2.1^2 \\ 2.1^2 & 4.9^2] \,,
\end{equation}
where both $S_v$ and $L$ are in units of MeV.
The Pearson correlation coefficient is $\rho \approx 0.98$.
As one might expect, these chiral NN and 3N interactions predict $(S_v,L)$ much lower (and more accurately) than the PREX--II-informed result obtained in Ref.~\cite{Reed:2021nqk}, $S_v = 38.1 \pm 4.7 \MeV$ and $L = 106 \pm 37 \MeV$ at the $1\sigma$ level.
For the incompressibility, we find $K \approx 235 \pm 5 \MeV$ at the 95\% credibility level, consistent with the prior distribution, as expected.

\begin{figure}[tb]
    \centering
    \includegraphics{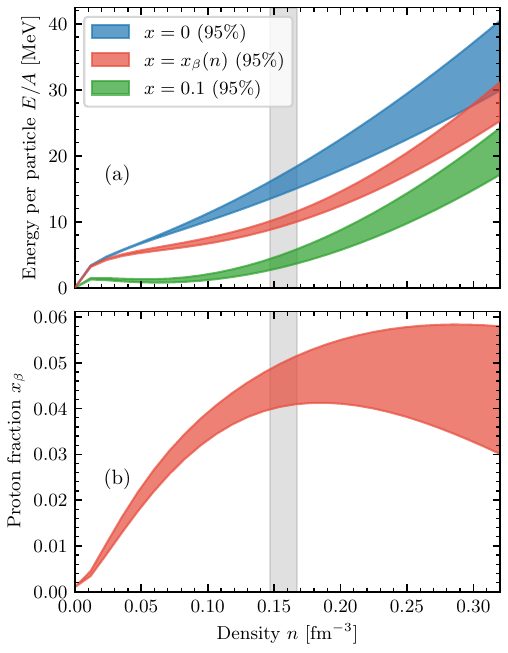}
    \caption{%
    Results for neutron-rich matter based on the EOS model~\eqref{eq:eos_model}. 
    Panel~(a) shows the energy per particle for matter with $x=0$ (i.e., PNM; blue band) and $x=0.1$ (green band) in addition to charge-neutral, $\beta$-equilibrated (neutron star) matter (red band). 
    All uncertainty bands depict the 95\% confidence region. 
    Panel~(b) depicts the proton fraction associated with $\beta$-equilibrated (neutron star) matter. 
    At the 95\% confidence level, the proton fraction of neutron star matter does not exceed $\approx 6\%$ up to twice the saturation density.
    The gray vertical bands represent the 95\% confidence interval for the empirical saturation density. %
    }
    \label{fig:nsm}
\end{figure}

Finally, we study the EOS and composition of ($\beta$-equilibrated) neutron star matter. 
At a given density $n$, we determine the proton fraction associated with charge-neutral $\beta$-equilibrium, $x_\beta(n)$, by solving the nonlinear root-finding problem~\cite{Hebeler:2013nza}
\begin{equation} \label{eq:beta_equilib}
\left[ \frac{\partial}{\partial x}\bar{\varepsilon}(n, x) + \mu_e(n,x) - \Delta m \right]_{x = x_\beta(n)} = 0 \,,
\end{equation}
with the chemical potential of the ultra-relativistic electrons,  $\mu_e(n, x) = \sqrt[3]{3\pi^2 x n}$, and the nucleon mass difference, $\Delta m = m_n - m_p \approx 1.29 \MeV$.
Figure~\ref{fig:nsm} shows our results for the neutron-rich matter EOS based on the parametric model~\eqref{eq:eos_model}. 
In Fig.~\ref{fig:nsm}(a), the energy per particle is shown for matter with $x=0$ (blue band; PNM) and $x=0.1$ (green band), both serving as references, in addition to neutron star matter with $x(n)=x_\beta(n)$ (red band). 
All uncertainty bands encompass the 95\% confidence region. 
In Fig.~\ref{fig:nsm}(b), the proton fraction density dependence of the proton fraction associated with $\beta$-equilibrium is shown. 
At the 95\% confidence level, the proton fraction of neutron star matter does not exceed $\approx 6\%$ up to twice the saturation density, consistent with our findings in Ref.~\cite{Drischler:2020fvz}.
The gray vertical bands represent the 95\% confidence level for the empirical saturation density, which serves as the reference density in this context.
Both panels use the same color coding.
We find these constraints at the 95\% confidence level:
\begin{subequations}
    \begin{align}
        \frac{E}{A} \left(n_0^\mathrm{(emp)}, x_\beta(n_0^\mathrm{(emp)})\right) & \approx 10.1 \pm 1.0 \MeV \,,\\ 
        x_\beta\left(n_0^\mathrm{(emp)}\right) & \approx 0.045 \pm 0.005 \,.
    \end{align}
\end{subequations}
Samples of the EOS and derived quantities, including $x_\beta(n)$, can be readily obtained from our publicly available source code~\cite{BUQEYEsoftware}.

\section{\rev{Explicit asymmetric nuclear matter calculations using MBPT(4)}}
\label{sec:anm}

\begin{figure*}[tb]
    \centering
    \includegraphics[width=\textwidth]{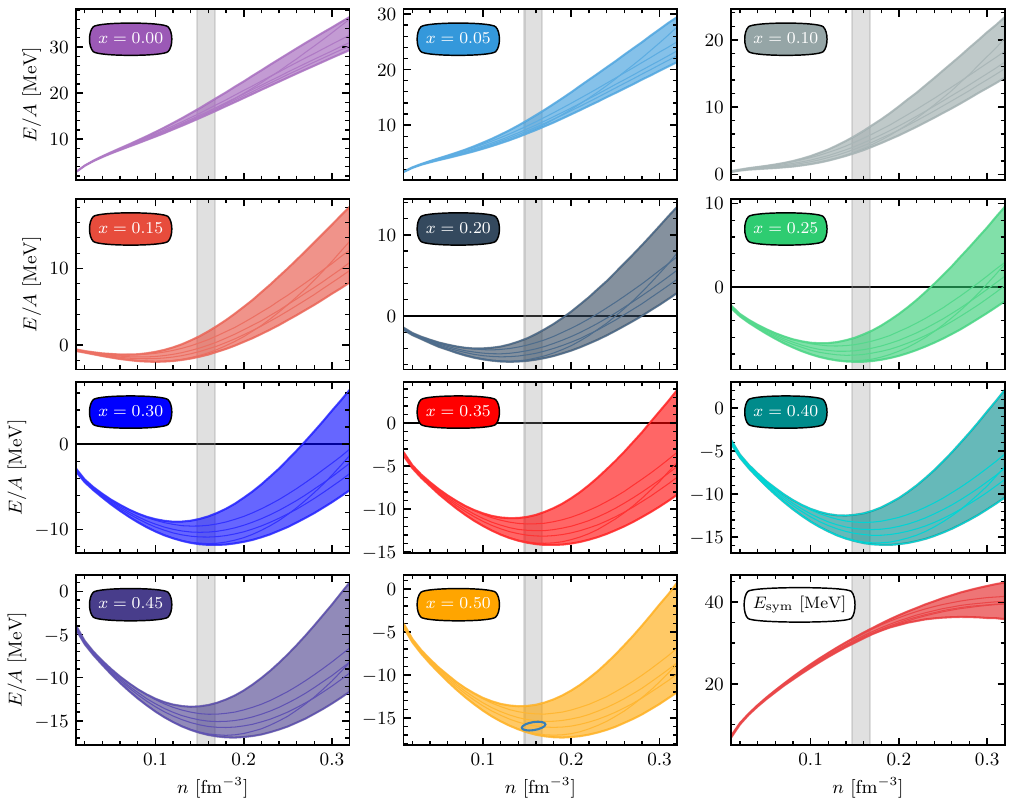}
    \caption{%
    Energy per particle of asymmetric matter as a function of the density $n$, sampled at 11 equidistant proton fractions $x$ between PNM ($x=0$) and SNM ($x=0.5$).  
    All results are based on our MBPT calculations up to fourth (third) order at the normal-ordered two-body (three-body) level.
    The panel with label ``$E_\text{sym}$'' (bottom-right panel) depicts the nuclear symmetry energy~\eqref{eq:esym}.
    The empirical range for the saturation density is highlighted in all panels by the vertical band.
    As in Fig.~\ref{fig:eos_pnm_snm_5th}, the blue ellipse in the panel for SNM ($x=0.5$) encompasses the 95\% credibility region of the empirical saturation point~\eqref{eq:emp_sat_point}. %
    }
    \label{fig:eos}
\end{figure*}

We perform explicit asymmetric matter calculations for $n \leqslant 0.32 \fmiq$ using MBPT(4), including residual 3N contributions up to third order, based on the six Hebeler~\etal\ interactions. 
The MBPT expansion is truncated at fourth order because of the rapid many-body convergence observed in Sec.~\ref{sec:mb_conv_res_heb} for these interactions and the computationally expensive nature of higher-order MBPT calculations.

\subsection{\rev{EOS constraints}}

Figure~\ref{fig:eos} shows the resulting energies per particle as a function of~$n$, sampled at 11 equidistant proton fractions~$x$ between PNM ($x=0$) and SNM ($x=0.5$).
In addition, the bottom-right panel (labeled ``$E_\text{sym}$'') depicts the symmetry energy~\eqref{eq:esym} in the standard quadratic approximation, which we find to be very accurate for these interactions (see also below).
The vertical bands highlight the empirical range of the saturation density, and in the SNM panel, the blue 95\% confidence ellipse depicts the empirical saturation point corresponding to the one in Fig.~\ref{fig:coester_plot}.
The colored shaded bands indicate, as before in Sec.~\ref{sec:pnm_snm_mbpt5}, the range of the energy per particle predicted by these interactions (depicted by the six lines).

As can be seen in Fig.~\ref{fig:eos}, the EOS of neutron-rich matter is tightly constrained at $n \lesssim n_0$, in contrast to matter closer to SNM \cite{Drischler:2015eba}. 
The spread among the six Hamiltonians increases with both density and proton fraction due to growing 3N contributions. 
For example, at $n = 0.16,\fmiq$, we obtain the ranges $E/A = 15.4-17.9 \MeV$, $-(8.9-5.6) \MeV$, and $-(17.0-13.5) \MeV$ for $x = 0.0$, $0.25$, and $0.50$, respectively.
At $n \gtrsim n_0$, 3N forces generally contribute substantially to the energy per particle and dominate the theoretical uncertainty~\cite{Hebeler:2020ocj,Drischler:2021kxf}. 
These forces are simpler in PNM than in matter with $x > 0$, since the short- and intermediate-range 3N terms proportional to $c_D$ and $c_E$ vanish~\cite{Hebeler:2009iv}. 
As a result, uncertainties are significantly reduced in PNM.
For $x \gtrsim 0.15$, we find that nuclear matter becomes bound below $n_0$, and saturation properties begin to emerge.

\subsection{\rev{Global parametric EOS models}}

To disseminate these asymmetric matter calculations, we construct global parametric EOS models using symbolic regression (SR), an emerging machine learning method~\cite{Angelis:2023review,Makke2024review}.
SR is a data-driven modeling method that identifies explicit mathematical expressions to describe data (here, $E(\delta,n)/A$) without relying on a priori known functional forms.
It explores combinations of mathematical operators and variables to identify equations that balance predictive accuracy with computational simplicity. 
Further details on SR are provided in the recent review articles~\cite{Angelis:2023review,Makke2024review} and their references, while recent applications are discussed in Refs.~\cite{Udrescu:2021,Keren:2023,Bakurov:2024rmu,Ghorashi:2025,Ghorashi:2025}.
We emphasize that, in this initial study, our goal is to explore SR’s ability to learn useful parametric EOS models from microscopic EOS calculations at low densities, rather than to provide a comprehensive investigation of its application, which we leave for future work. 

We use the Python package \texttt{PySR}~\cite{cranmerInterpretableMachineLearning2023} to perform SR on the results presented in Fig.~\ref{fig:eos} (see also Refs.~\cite{Udrescu:2019mnk,Udrescu2020,NEURIPS2020:33a854e2} for the AI Feynman package).
The \texttt{PySR} regressor is configured for evolutionary search, running up to 3000 iterations with a population of 1200 expressions, each evolved over 400 cycles per iteration.
We limit the model complexity to a maximum size of 32 and depth of 12, and the total runtime to 4~h. 
The search space is restricted to the binary operations of addition and multiplication involving the Fermi momentum  $k_\text{F} \propto \sqrt[3]{n}$ (in PNM and SNM) and $\delta = 1-2x$, producing only polynomial expressions in those variables to avoid unphysical poles in the low-density EOS. 
We deliberately do not assume that the EOS depends approximately quadratically on $\delta$.

For a given interaction, we train parametric models within this search space to the scaled energy per particle, $E(\delta,n/n_0^\star)/(16 \MeV \, A)$, as a function of $\delta \leqslant 1$ and the scaled density, $\bar{n} = n/n_0^\star \leqslant 2$, using the standard mean-squared error loss function, analogous to least-squares fits. 
Both the scaled energy per particle and scaled density are constructed to be dimensionless and of order one or less to improve training stability.
This procedure leads to parametric models of the form of Eq.~(18) in Ref.~\cite{Drischler:2015eba}:
\begin{equation} \label{eq:e-a-semiana}
    \frac{E}{A}(\delta,\bar{n}) = 16 \MeV \; \sum \limits_{\mu,\nu\rev{\geqslant}0} C_{\mu\nu} \; \delta^\mu \; \bar{n}^\frac{\nu}{3} \,,
\end{equation}
with the to-be-determined model parameters $C_{\mu\nu}$.
However, whereas Ref.~\cite{Drischler:2015eba} fixed the exponents to $\mu = 0,2$ and $\nu = 2,3\ldots,6$ (i.e., 10 terms in total), we instead use \texttt{PySR} to determine suitable values for them.

\begin{table*}[tb]
\renewcommand{\arraystretch}{1.15}
\caption{%
Coefficients $C_{\mu\nu}$ of the global EOS model~\eqref{eq:e-a-semiana}, calibrated to the calculated $E(\delta,n)/A$ for each of the six interactions listed in Table~\ref{tab:hebeler_interactions} individually. 
The coefficients are dimensionless.
All other coefficients in Eq.~\eqref{eq:e-a-semiana} are zero. %
}
\label{tab:fit_coeffs}
\begin{ruledtabular}
\begin{tabular}{ld{1.4}d{1.4}d{1.4}d{1.4}d{1.4}d{1.4}d{1.4}d{1.4}}
\multicolumn{1}{c}{Interaction}& \multicolumn{1}{c}{$C_{01}$} & \multicolumn{1}{c}{$C_{03}$} & \multicolumn{1}{c}{$C_{04}$} & \multicolumn{1}{c}{$C_{06}$} & \multicolumn{1}{c}{$C_{09}$} & \multicolumn{1}{c}{$C_{21}$} & \multicolumn{1}{c}{$C_{23}$} & \multicolumn{1}{c}{$C_{26}$} \\ 
\hline
(1.8/2.0)  & -0.6808 & 0.9576 & -2.8837 & 1.7292 & -0.1832 & 0.8926 & 1.5146 & -0.3898 \\
(2.0/2.0)  & -0.6689 & 0.8907 & -2.7322 & 1.6948 & -0.1819 & 0.8963 & 1.4734 & -0.3905 \\
(2.0/2.5)   &-0.6972 & 1.1585 & -3.0822 & 1.6867 & -0.1194 & 0.9336 & 1.3794 & -0.2809 \\
(2.2/2.0)  & -0.6545 & 0.8089 & -2.5939 & 1.6626 & -0.1808 & 0.8953 & 1.4532 & -0.3892 \\
(2.8/2.0)  & -0.6056 & 0.5166 & -2.1984 & 1.5555 & -0.1733 & 0.8856 & 1.4346 & -0.3786 \\
(2.0/2.0)*  & -0.7133 & 1.7665 & -4.1863 & 2.5990 & -0.3051 & 0.8296 & 1.6436 & -0.5244
\end{tabular}
\end{ruledtabular}
\end{table*}

While it would be possible to determine separate (and likely locally more accurate) parametrizations of the form given in Eq.~\eqref{eq:e-a-semiana} for each of the Hebeler~\etal\ interactions, we see an advantage in adopting a single parametrization that describes them all.
Our non-exhaustive exploration of the search space with \texttt{PySR} suggests that $\mu = 1, 3, 4, 6, 9$ for $\nu = 0$ and $\mu = 1, 3, 6$ for $\nu = 2$ (i.e., 8 terms in total) is a reasonable compromise between accuracy and computational complexity (see below).
The corresponding values for the coefficients $C_{\mu\nu}$ are summarized in Table~\ref{tab:fit_coeffs}.

A few comments are in order:
\texttt{PySR} found on its own the standard quadratic approximation for the asymmetry dependence of the EOS. 
The five $C_{\mu,\nu =0}$ coefficients describe the density dependence of $E(\delta=0,\bar{n})/A$, while the three $C_{\mu,\nu =2}$ coefficients describe the density dependence of $E_{\text{sym}}(\bar{n})$.
As shown in Fig.~\ref{fig:eos}, the density dependence of $E_{\text{sym}}(\bar{n})$ is simpler, as it increases monotonically, whereas $E(\delta=0, \bar{n})/A$ exhibits nuclear saturation.
Hence, it may not be surprising that the latter requires more terms in the expansion to achieve sufficient accuracy.
We have verified that the resulting parametric model accurately reproduces the underlying microscopic calculations, including those in the neutron-rich regime. 
At $n \leqslant 2n_0^\star$, the median and standard deviation of the maximum errors across the six interactions are $192 \pm 85 \keV$ in PNM and $245 \pm 48 \keV$ in SNM, so similar to the accuracy reported in Ref.~\cite{Drischler:2015eba}, even though that study included two additional terms in Eq.~\eqref{eq:e-a-semiana} and considered only $n \lesssim 1.4 n_0^\star$.
In contrast to Ref.~\cite[see Table~II]{Drischler:2015eba}, which showed clear signs of overfitting, manifested by large coefficients (i.e., \(|C_{\mu\nu}| \lesssim 27\) in units of $16 \MeV$) and substantial cancellations, we find that the extracted coefficients are much better behaved (i.e., of order one).
Nevertheless, some cancellations still remain due to coefficients with alternating signs, as seen in Table~\ref{tab:fit_coeffs}, which may be mitigated by enlarging the search space.
In future work, it will be interesting to apply SR more systematically to microscopic EOS calculations to develop refined functional forms and assess their robustness in the neutron-rich regime and across different nuclear interactions.

\section{\rev{Many-body convergence of the DHS interactions up to MBPT(4)}} 
\label{sec:fit_interactions}

The second set of interactions we study were developed in Ref.~\cite{Drischler:2017wtt} by combining the EMN NN potentials~\cite{Entem:2017gor} with 3N
forces at the same chiral order (i.e., \NNLO and \NNNLO) and momentum cutoff.
The 3N LECs $c_D$ and $c_E$ were adjusted to the triton binding energy and, to the extent possible, the empirical saturation point in SNM. 
For two momentum cutoffs, Ref.~\cite{Drischler:2017wtt} obtained three 3N forces with different combinations of $c_D$ and $c_E$ and reasonable saturation properties at \NNLO and \NNNLO. 
As in Refs.~\cite{Drischler:2020yad,Drischler:2020hwi}, we consider here one of these $(c_D,c_E)$ combinations at \NNLO ($c_D=-1.75$, $c_E=-0.64$) and one at \NNNLO ($c_D=-3.00$, $c_E=-2.22$), respectively, both with the larger 3N momentum cutoff, $\Lambda_\text{3N} = 500 \MeV$.
We choose these interactions because they are unevolved in contrast to the Hebeler~\etal\ potentials and are available up to $\NkLO{3}$, including subleading 3N forces.
The MBPT convergence is therefore expected to be slower for these bare interactions than before. 
Furthermore, this choice allows us to compare the size of MBPT contributions to the EFT truncation errors quantified by the BUQEYE collaboration in Refs.~\cite{Drischler:2020yad,Drischler:2020hwi} based on the EOS calculations in Ref.~\cite{Drischler:2017wtt}.

\begin{figure}[tb]
    \centering
    \includegraphics[width=\columnwidth]{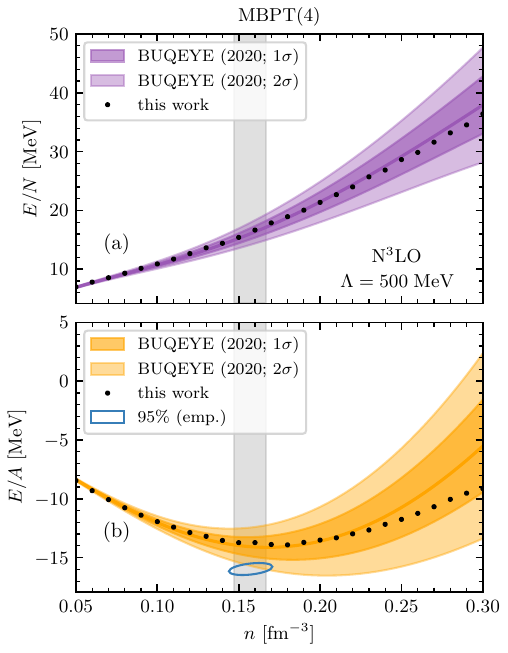}
    \caption{%
    Our MBPT(4) results for the DHS interactions at \NkLO{3} for the momentum cutoff $\Lambda_\mathrm{3N} = 500 \MeV$ in PNM (panel~a) and SNM (panel~b) are depicted by the black dots (``this work'').
    The figure overlays our results with the uncertainty bands reported by the BUQEYE collaboration~\cite{Drischler:2020yad,Drischler:2020hwi} for estimating the EFT truncation error at two confidence levels: 68\% (darker shading) and 95\% (lighter shading).  
    As in Fig.~\ref{fig:eos_pnm_snm_5th}, the blue ellipse in the lower panel encompasses the 95\% credibility region of the empirical saturation point~\eqref{eq:emp_sat_point}. %
    }
    \label{fig:eos_buqeye_comparison}
\end{figure}

\begin{table*}[tb]
\renewcommand{\arraystretch}{1.1}
\caption{
A sample of the order-by-order MBPT convergence for the DHS NN and 3N interactions at \NNNLO (top) and \NNLO (bottom), and momentum cutoff $\Lambda_\text{3N} =500 \MeV$.
The values of the 3N LECs are annotated.
The MBPT contributions up to fourth order, including separate columns for second-order (``2nd (3b)'') and third-order (``3rd (3b)'') residual 3N contributions, are given at $n = 0.08$, $0.16$, and $0.30 \fmiq$ in units of MeV.
The total contribution from the second (third) order is then given by the sum of the two-body (``2b'') and three-body (``3b'') contributions at each order, i.e., the columns labeled ``2nd (2b)'' and ``2nd (3b)'' (``3rd (2b)'' and ``3rd (3b)''), corresponding to the columns ``2nd'' (``3rd'') in Table~\ref{tab:eos}.
For the short-hand notation, see the caption of Table~\ref{tab:eos}.%
}
\label{tab:fit_eos}
\begin{tabular*}{\textwidth}{l}
\vspace{-3pt} \\
\NNNLO DHS (EMN~500~MeV), with $c_D=-3.00$ and $c_E=-2.22$ \\
\end{tabular*}

\vspace{2pt} \begin{ruledtabular}
\begin{tabular}{lccccccccc} {PD} & {HF} & {2nd (2b)}&{2nd (3b)}& {3rd (2b)}&{3rd (3b)} & {4th} &  {Total} & {Pad{\'e}[2/2]}  \\
\hline \\[-1.5ex]
P08 & \phantom{-}12.64(00) & -3.92(01) & \phantom{0}-0.05(01) & \phantom{0}\phantom{-}0.57(01) & \phantom{0}\phantom{-}0.19(03) & \phantom{0}-0.11(03) & \phantom{0}\phantom{-}9.33(04) & \phantom{0}\phantom{-}9.34 \\
P16 & \phantom{-}23.67(00) & -7.95(01) & \phantom{0}-0.08(01) & \phantom{0}\phantom{-}0.82(01) & \phantom{0}\phantom{-}0.26(05) & \phantom{0}-0.03(05) & \phantom{-}16.68(08) & \phantom{-}16.67 \\
P30 & \phantom{-}50.76(00) & -13.14(02) & \phantom{0}-0.04(01) & \phantom{0}\phantom{-}0.50(04) & \phantom{0}-0.10(09) & -1.55(09) & \phantom{-}36.42(13) & \phantom{-}36.32 \\
\hline
S08 & -1.42(00) & -8.97(01) & \phantom{0}-0.32(01) & \phantom{0}-0.36(01) & \phantom{0}-0.40(01) & \phantom{0}\phantom{-}0.72(04) & -10.75(04) & -10.69 \\
S16 & \phantom{-}1.71(00) & -14.63(01) & \phantom{0}-0.84(01) & \phantom{0}-0.99(02) & \phantom{0}-0.08(04) & \phantom{-}1.08(06) & -13.76(08) & -13.69 \\
S30 & \phantom{-}21.48(00) & -25.43(02) & -1.38(01) & \phantom{0}-0.83(04) & \phantom{0}\phantom{-}0.22(10) & -3.16(08) & \phantom{0}-9.11(13) & \phantom{0}-9.71 \\
\end{tabular}
\end{ruledtabular}
\begin{tabular*}{\textwidth}{l}
\vspace{-3pt} \\
\NNLO DHS (EMN~500~MeV), with $c_D=-1.75$ and $c_E=-0.64$ \\
\end{tabular*}

\vspace{2pt} \begin{ruledtabular}
\begin{tabular}{lccccccccc} {PD} & {HF} & {2nd (2b)}&{2nd (3b)}& {3rd (2b)}&{3rd (3b)} & {4th} &  {Total} & {Pad{\'e}[2/2]}  \\
\hline \\[-1.5ex]
P08 & \phantom{-}11.35(00) & -2.04(01) & \phantom{0}-0.03(01) & \phantom{0}\phantom{-}0.38(01) & \phantom{0}\phantom{-}0.06(01) & \phantom{0}-0.15(01) & \phantom{0}\phantom{-}9.56(01) & \phantom{0}\phantom{-}9.59 \\
P16 & \phantom{-}21.14(00) & -4.07(01) & \phantom{0}-0.04(01) & \phantom{0}\phantom{-}0.51(01) & \phantom{0}\phantom{-}0.09(02) & \phantom{0}-0.30(02) & \phantom{-}17.32(03) & \phantom{-}17.37 \\
P30 & \phantom{-}45.81(00) & -6.38(01) & \phantom{0}-0.02(01) & \phantom{0}\phantom{-}0.93(02) & \phantom{0}-0.14(03) & \phantom{0}-0.93(04) & \phantom{-}39.28(05) & \phantom{-}39.38 \\
\hline
S08 & -1.82(00) & -8.74(01) & \phantom{0}-0.30(01) & \phantom{0}\phantom{-}0.22(01) & -1.18(01) & \phantom{0}\phantom{-}0.63(01) & -11.20(02) & -11.10 \\
S16 & \phantom{-}1.21(00) & -14.02(02) & \phantom{0}-0.64(01) & \phantom{0}-0.66(01) & \phantom{0}-0.65(03) & \phantom{0}\phantom{-}0.48(02) & -14.29(04) & -14.21 \\
S30 & \phantom{-}24.09(00) & -24.66(02) & \phantom{0}-0.80(01) & \phantom{0}\phantom{-}0.93(03) & \phantom{-}1.07(05) & -7.70(04) & \phantom{0}-7.06(06) & \phantom{0}-8.45 
\end{tabular}
\end{ruledtabular}
\end{table*}

The black dots in Fig.~\ref{fig:eos_buqeye_comparison} depict our MBPT(4) results for energy per particle based on the \NNNLO DHS interaction in PNM [Fig.~\ref{fig:eos_buqeye_comparison}(a)] and SNM [Fig.~\ref{fig:eos_buqeye_comparison}(b)].
We obtain qualitatively similar results for the \NNLO potential and thus provide them in Appendix~\ref{app:add_results} for completeness.
Given our available computational resources and the high cost of normal ordering \NNNLO 3N forces \rev{due to their rich analytical structure~\cite{Bernard:2007sp,Bernard:2011zr,Hebeler:2020ocj}}, we have been able to carry out MBPT calculations only up to fourth order, i.e., MBPT(4).
That means, compared with Ref.~\cite{Drischler:2017wtt}, normal-ordered 3N contributions at fourth order and residual 3N contributions at third order are included.
In addition, Table~\ref{tab:fit_eos} summarizes the associated order-by-order MBPT results:
The upper part lists the \NNNLO results, while the lower part shows those at \NNLO, and the corresponding 3N LECs are indicated in each case.
The individual MBPT contributions up to fourth order are reported at $n = 0.08$, $0.16$, and $0.30 \fmiq$ in MeV. 
Separate columns are provided for the residual 3N contributions at second order (``2nd (3b)'') and third order (``3rd (3b)''). 
The total contribution at second (third) order is obtained by summing the two- and three-body terms.
That is, the sum of the columns labeled ``2nd (2b)'' and ``2nd (3b)'' (``3rd (2b)'' and ``3rd (3b)''), which then corresponds to the ``2nd'' (``3rd'') columns in Table~\ref{tab:eos}. 
Since the MBPT results are available up to fourth order, we also provide the results of the symmetric Pad{\'e}[2/2].
Tables~\ref{tab:eos} and~\ref{tab:fit_eos} use the shorthand notation.

To put the additional MBPT contributions in perspective, Fig.~\ref{fig:eos_buqeye_comparison} compares our results (black dots) with the correlated EFT truncation-error uncertainty bands at the $1\sigma$ (68\%; darker shading) and $2\sigma$ (95\%; lighter shading) confidence levels, obtained from the same nuclear interactions, as reported by the BUQEYE collaboration~\cite{Drischler:2020yad,Drischler:2020hwi}.
Our calculations (``this work'') incorporate the normal-ordered 3N terms at fourth order as well as the residual 3N terms at third order, which were not considered in Refs.~\cite{Drischler:2020hwi,Drischler:2020yad}.
With these additional contributions, we find that the energy per particle remains well within these $1\sigma$ EFT truncation-error bands across the entire density range shown, particularly in PNM at $n \lesssim 2 n_0$ and SNM at $n \lesssim 1.5 n_0$.

However, as shown in Table~\ref{tab:fit_eos}, we do not generally observe the convergence pattern found in Sec.~\ref{sec:mb_conv_res_heb} for the softer Hebeler~\etal\ interactions, where the difference between successive MBPT orders decreases mostly monotonically, except for PNM at $n \lesssim n_0$. 
Fourth-order contributions are suppressed relative to those at second order, but not necessarily relative to third-order contributions. 
Furthermore, the third-order residual 3N contributions are not systematically suppressed compared to those at second order or to the normal-ordered 3N contributions. 
Hence, it would be interesting to study the fourth-order residual 3N contributions in future work to determine whether these patterns persist.

On the other hand, these additional contributions result in more net attraction in SNM at $n \gtrsim 1.5 n_0$, thereby markedly altering the density dependence of the EOS in this regime, driving the energy per particle close to the lower $1\sigma$ boundary of the EFT truncation error.
This finding is also reflected by the large attractive fourth-order contributions reported in Table~\ref{tab:fit_eos}, as also indicated by the MeV-level discrepancies between the total energy and the Pad{\'e}[2/2] approximant at the highest density and both chiral orders.
For $n \lesssim 1.5 n_0$, the net impact of the additional contributions on the EOS is not significant, and thus they do not improve the saturation properties of the potentials relative to the empirical point (i.e., the blue ellipse in Fig.~\ref{fig:eos_buqeye_comparison}(b)).
In this density region, the Pad{\'e}[2/2] approximants and the MBPT estimates of the total energy are consistent with one another at or below the $100 \keV$ level.

Overall, these results suggest reasonable many-body convergence in PNM at $n \lesssim 2 n_0$ and in SNM at $n \lesssim 1.5 n_0$ for these unevolved NN and 3N interactions.
At higher densities in SNM, however, our results suggest a possible breakdown of the MBPT expansion, which warrants further investigation, e.g., through benchmarks against nonperturbative many-body frameworks.

\section{Summary and Outlook} 
\label{sec:summary_outlook}

We have presented an automated, GPU-accelerated framework for MBPT calculations of the zero-temperature EOS based on chiral NN and 3N forces, extending previous work~\cite{Drischler:2017wtt} to fifth (and even sixth) order at the normal-ordered two-body level. 
Combining automated diagram generation and evaluation, all MBPT diagrams have been implemented up to sixth order at the normal-ordered two-body level and evaluated up to fifth order. 
Residual three-body diagrams have been computed up to third order~\cite{Hu:2018dza}, thereby completing our MBPT calculations at this order.
We are not aware of other MBPT calculations based on chiral NN and 3N forces at these high MBPT orders.
This work represents a significant advance in the automation and exploration of MBPT for nuclear matter, enabling improved calculations of the EOS as a function of the density and proton fraction with controlled numerical uncertainties~\cite{Drischler:2021kxf}.
Moreover, the observed cancellations among diagrammatic contributions reinforce earlier findings~\cite{Holt:2016pjb}, indicating that, at a given MBPT order and many-body level, diagrams should be evaluated consistently as a complete set rather than selectively included.
Although we have implemented MBPT(6) in preparation for future studies, more computational resources than we have currently available would be necessary to evaluate it in practice.

Specifically, we used the automated diagram generator developed by Arthuis~\etal~\cite{Arthuis:2018yoo,Arthuis:2020tjz,Tichai:2021ewr} to derive the MBPT diagrams and associated expressions up to sixth order at the two-body level.
These expressions are summarized in Table~\ref{tab:mbpt} and are also provided in a machine-readable format in the accompanying GitHub repository~\cite{Drischler_Cheat_sheets}, along with the residual 3N contributions up to third order.
For example, MBPT(5) and MBPT(6) yield 840 and 27300 diagrams at the two-body level, respectively, making both the implementation and accurate evaluation of them computationally challenging. 
Building on prior work~\cite{Drischler:2017wtt}, we developed a hybrid strategy for automated diagram evaluation that combines diagram-specific source code generation with template-based runtime optimization, enabling efficient computation of partial sums of these diagrams.
This strategy is crucial for handling removable divergences in individual MBPT diagrams~\cite{Wellenhofer:2021eis} order by order and for improving overall numerical accuracy. 

Several technical advances were necessary to enable our high-order MBPT calculation. 
We used GPU acceleration to mitigate the computational cost of normal-ordering 3N forces without relying on common approximations, such as zero total momentum or angular averaging of the effective two-body potentials~\cite{Hebeler:2009iv,Hebeler:2010xb,Drischler:2015eba,Holt:2019bah,Drischler:2021kxf} (see also Ref.~\cite{Djarv:2021xjg}). 
While CPUs evaluate the NN interactions, one or multiple GPUs normal order the 3N forces on the fly, resulting in computational speedups that exceed two orders of magnitude relative to CPU-only implementations. 
This hybrid CPU-GPU approach allowed us to include normal-ordered 3N contributions at fourth (and fifth) order in contrast to, e.g., Ref.~\cite{Drischler:2017wtt}.
In addition, the spin-isospin traces, which scale exponentially with MBPT order, have been significantly optimized through structural analysis of the diagrams and GPU acceleration.
Furthermore, we developed \pvegas, a high-performance MC integrator that utilizes adaptive importance sampling and supports vector-valued integrands, to efficiently and accurately evaluate high-dimensional momentum integrals (see Table~\ref{tab:mbpt}). 
Finally, the thousands of computational tasks necessary to map the EOS's dependence on the density and proton fraction at high MBPT orders are orchestrated by \mpijm~\cite{Berkowitz:2017xna,Berkowitz:2018gqe}, a custom job manager that enables efficient scheduling and fault-tolerant execution on leadership-class supercomputers. 
Together, these advances allow for large-scale, high-order MBPT calculations with controlled numerical uncertainties and reproducible results.

We then studied the MBPT convergence up to fifth order (``MBPT(5)'') up to about twice the saturation density based on two sets of chiral NN and 3N interactions: 
the SRG-evolved Hebeler~\etal\ interactions~\cite{Hebeler:2010xb} and the unevolved \NNLO and \NNNLO DHS potentials~\cite{Drischler:2017wtt}. 
In particular, for the soft Hebeler~\etal\ interactions in PNM, we observe rapid MBPT convergence, with the differences between successive MBPT orders generally decreasing monotonically (see Table~\ref{tab:eos}).
Fifth-order contributions are essentially negligible in light of the total energy per particle and the expected EFT truncation error (of several MeV). 
As expected, the convergence for the same interactions in SNM is slightly slower than that in PNM but still controlled, especially for $n \leqslant 0.16 \fmiq$. 
Hence, we observed a similarly good convergence pattern for the predicted saturation point, but fifth-order contributions could not reconcile the known disagreement between the predicted and empirical saturation points for these chiral interactions. 
At $n \geqslant 0.16 \fmiq$, we found fifth-order contributions of the order of only $100 \keV$ for the hardest interaction. 
We concluded that these potentials are perturbative in the density region studied here, owing to the SRG evolution to lower resolution scales.

For the harder, more computationally demanding DHS interactions, we were able to study the MBPT convergence pattern only up to fourth order.
Updating the MBPT calculations in Ref.~\cite{Drischler:2017wtt} to include fourth-order normal-ordered 3N and third-order residual 3N contributions, we find that the energy per particle generally remains within the $1\sigma$ EFT truncation-error bands in PNM and SNM at $n \lesssim 2n_0$~\cite{Drischler:2020hwi,Drischler:2020yad}. 
However, the MBPT contributions do not always decrease monotonically with increasing order, and the third- and fourth-order terms are not systematically suppressed.
In SNM at $n \gtrsim 1.5n_0$, the additional MBPT contributions are net-attractive, markedly altering the density dependence of the EOS in this regime and suggesting a possible breakdown of the MBPT expansion.
Overall, the results indicate reasonable many-body convergence at low to moderate densities, especially in PNM, while SNM at $n \gtrsim 1.5n_0$ warrants further nonperturbative benchmarks. 

Using our MBPT(5) calculations of the PNM EOS and the empirical saturation point, we calibrated a parametric model of asymmetric matter to study the EOS and composition of neutron star matter in the uniform phase while maintaining realistic saturation properties in SNM. 
We found that, consistent with previous work, e.g., in Ref.~\cite{Drischler:2020fvz}, the proton fraction of neutron star matter at the 95\% credibility level does not exceed 6\% at twice the saturation density.
Furthermore, we performed explicit calculations of the asymmetric-matter EOS and leveraged symbolic regression, a powerful machine-learning method, to develop new parametric models of low-density nuclear matter.

Looking ahead, the developed MBPT framework offers several avenues for future research that are both feasible and promising. 
These include systematic order-by-order studies of asymmetric nuclear matter for bare, more modern chiral interactions, such as the LENPIC SMS potentials~\cite{Reinert:2017usi,Epelbaum:2022cyo}, quantification of EFT truncation errors~\cite{Drischler:2020hwi,Drischler:2020yad}, and finite-temperature extension of the MBPT framework~\cite{Keller:2020qhx,Drischler:2021kxf,Keller:2022crb}. 
Of particular interest for clarifying many-body convergence across different SRG resolution scales are systematic comparisons of MBPT with nonperturbative methods, including the IMSRG~\cite{Zhen:2025gfy} and AFDMC~\cite{Armstrong:2025tza,Lonardoni:2019ypg}, complemented by a comprehensive Weinberg eigenvalue analysis~\cite{Hoppe:2017lok,Reinert:2017usi,Weinberg:1963zza} at finite density~\cite{Ramanan:2007bb,Ramanan:2013mua} that includes normal-ordered 3N forces.
In turn, MBPT can highlight the importance of specific diagram classes such as fourth-order quadruple-excitation diagrams, \rev{some of} which are underestimated by IMSRG(2)~\cite{Hergert:2015awm,Li:2026njw}.
It can also serve as a diagnostic for how the (IM)SRG suppresses nonperturbative behavior in different many-body channels.

Extending the automated diagram generator to the three-body level would allow one to validate the derivation of the third-order residual terms in Ref.~\cite{Hu:2018dza} and to study even higher-order residual terms by explicit calculation, thereby pushing consistent order-by-order MBPT calculations beyond third order.
Work in this direction is ongoing, and our preliminary analysis indicates the presence of 887 and 80500 residual 3N diagrams at fourth and fifth order, respectively.
Taking into account anomalous and complex-conjugate diagrams at fourth order, this leaves 496 diagrams to be computed, so the same order of magnitude as two-body diagrams at fifth order.
Even without this extension, the newly included third-order residual 3N diagrams warrant further investigation across a wider range of chiral NN and 3N interactions to assess their importance for nuclear-matter observables and neutron-star inference.

The high computational costs of sixth-order MBPT calculations could be mitigated using approximated normal-ordering methods, whose accuracy can (and should) be benchmarked against our quasi-exact normal-ordering calculations.
This approach may offer a practical compromise between accuracy and computational cost.
Another promising direction for additional computational speed-ups would be to reformulate our MBPT framework as a diagrammatic MC approach~\cite{Brolli:2025ehf}, in which both MBPT diagrams and momenta are randomly sampled.
Furthermore, reducing computational costs across all MBPT orders is critical for propagating uncertainties in the LECs, which arise from calibrating the forces to scattering phase shifts and experimental data, to the nuclear EOS.
This propagation requires computationally demanding MC sampling of the LEC posterior distributions and MBPT calculations for each sample.

Various approaches to keep the computational costs tractable have recently been proposed.
For example, normalizing flows have been shown to yield efficient importance-sampling estimators for the MC evaluation of high-dimensional momentum integrals in our MBPT framework~\cite{Brady:2021plj,Wen:2024shw}.
These estimators have also been shown to transfer well from one point in the EOS parameter space (e.g., density and LEC value) to neighboring points, enabling more efficient mappings of the EOS parameter space. 
Another promising approach is using parametric matrix models (PMM)~\cite{Cook:2024toj} as fast and accurate emulators for the LEC dependence of the nuclear EOS.
Motivated by the reduced basis method, PMMs offers a remarkably simple yet powerful non-intrusive approach to constructing many-body emulators for principled UQ, e.g., as recently demonstrated in Ref.~\cite{Armstrong:2025tza}. 
Finally, advanced resummation techniques~\cite{Wellenhofer:2020ylh, Svensson:2025jde, gsum}, including physics-informed emulators~\cite{Demol:2019yjt,Demol:2020mzd,Cook:2024toj}, may prove useful for accelerating the MBPT convergence (in addition to RG methods) and extracting converged results from slowly converging MBPT expansions.
Explicit high-order MBPT calculations using the developed framework will enable the development and benchmarking of these resummation techniques.

Together with RG methods, MBPT provides a fertile ground for studying the nuclear EOS and benchmarking nonperturbative frameworks based on diagrammatic expansions of the many-body Schr{\"o}dinger equation.
The developments presented here establish a robust and scalable foundation for explicit high-order MBPT calculations of nuclear matter, thereby enabling a wide range of applications in nuclear theory and astrophysics.
Our publicly available GitHub repositories~\cite{Drischler_Cheat_sheets,BUQEYEsoftware} will help facilitate these applications.

\begin{acknowledgments}
We are most grateful to R.~J. Furnstahl for fruitful discussions, and Tong Li, Bai-Shan Hu, and Xu Furong for sharing the expressions for the residual 3N diagrams at third order~\cite{Hu:2018dza}.
C.D.\ thanks the NSF Physics Frontier Center N3AS, the Physics Department of the University of California, Berkeley, as well as the Nuclear Science Division at Lawrence Berkeley National Laboratory, for the warm hospitality during extended research visits. 
This material is based upon work supported by the NSF under Award No.\ PHY-2339043 and the U.S.\ Department of Energy, Office of Science, Office of Nuclear Physics, under the FRIB Theory Alliance Award No.\ DE-SC0013617.
This work was also in part supported by the U.S.\ Department of Energy, the Office of Science, the Office of Nuclear Physics, and SciDAC under Awards DE-SC00046548 and DE-AC02-05CH11231. 
The \mpijm development by K.~M.~for this project was supported by NSF Award No.\ OAC-2311431.
P.~A.~was supported by the European Union under the Marie Skłodowska-Curie Grant Agreement No.~101152722. 
We used computational resources of the Oak Ridge Leadership Computing Facility at the Oak Ridge National Laboratory, which is supported by the Office of Science of the U.S.\ Department of Energy under Contract No.\ DE-AC05-00OR22725.
The following open-source libraries were used to generate the results in this work:
\texttt{corner}~\cite{corner},
\texttt{EspressoDB}~\cite{Chang:2019khk},
\texttt{GSL}~\cite{gsl},
\texttt{Jupyter}~\cite{jupyter},
\texttt{matplotlib}~\cite{Hunter:2007},
\texttt{NetworkX}~\cite{SciPyProceedings_11},
\texttt{numpy}~\cite{harris2020array},
\texttt{PyMC}~\cite{abrill-pla2023pymc}, 
\texttt{PySR}~\cite{cranmerInterpretableMachineLearning2023}, 
\texttt{scipy}~\cite{2020SciPy-NMeth},
and
\texttt{sympy}~\cite{sympy}.

Views and opinions expressed are however those of the author(s) only and do not necessarily reflect those of the European Union or the European Research Executive Agency (REA). 
Neither the European Union nor the granting authority can be held responsible for them.
\end{acknowledgments}

\section*{Data availability}

The data that support the findings of this article are openly available~\cite{BUQEYEsoftware}.

\appendix
\section{Additional results}
\label{app:add_results}

\begin{figure}[tb]
    \begin{center}
    \includegraphics[width=\linewidth]{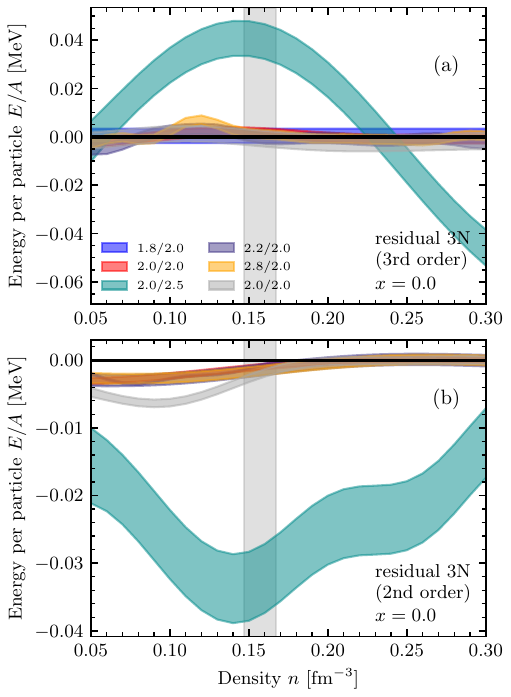}
    \end{center}
    \caption{%
    Residual 3N contributions to the energy per particle $E/A$ at third [panel~(a)] and second order [panel~(b)] in PNM for the six Hebeler~\etal\ interactions similar to Fig.~\ref{fig:residual3N_3rd_snm}.
    The labels in the legend indicate $\lambda_\text{SRG}$ and $\Lambda_\text{3N}$ (see also Table~\ref{tab:hebeler_interactions}).
    }
    \label{fig:residual3N_3rd_pnm}
\end{figure}

Figure~\ref{fig:residual3N_3rd_pnm} shows the residual 3N contributions to the energy per particle $E/A$ at third [Fig.~\ref{fig:residual3N_3rd_pnm}(a)] and second order [Fig.~\ref{fig:residual3N_3rd_pnm}(b)] for the six Hebeler~\etal\ potentials. 
This figure is analogous to Fig.~\ref{fig:residual3N_3rd_snm}, but for PNM. The (in magnitude)  largest contributions arise for the interaction with the larger cutoff, $\Lambda_\text{3N} = 2.5 \fmi$, as expected. 
However, they are about an order of magnitude smaller than the corresponding values in SNM. 
With $|E/A| \ll 40 \keV$ for most interactions across all densities shown, these contributions are negligible compared with the normal-ordered 3N contributions.

\begin{figure*}[tb]
    \centering
    \includegraphics[width=\textwidth]{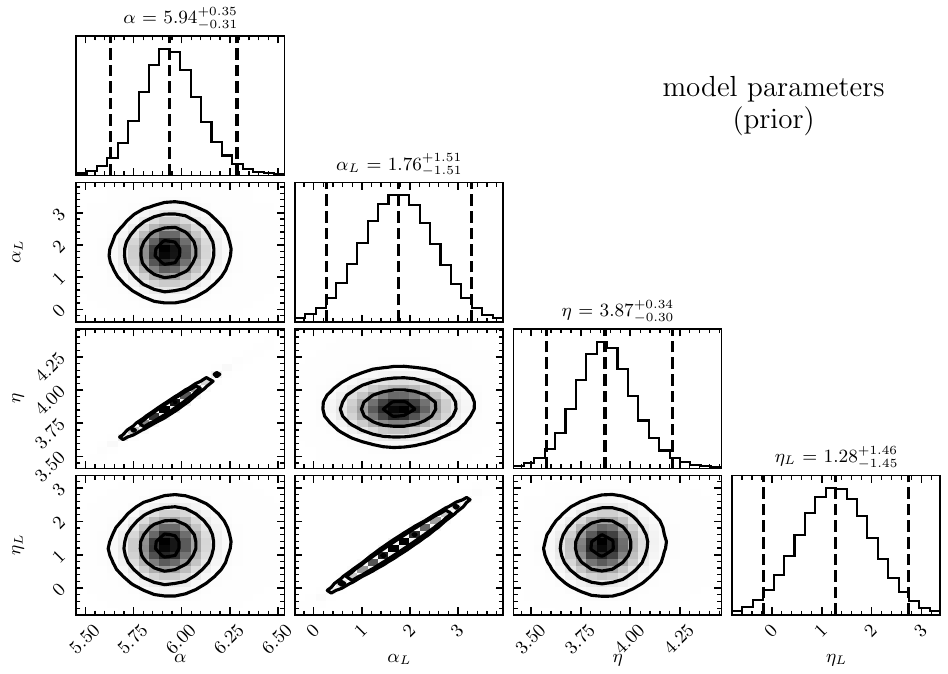}
    \caption{%
    Sampled prior distribution function for the parameters of the EOS model~\eqref{eq:eos_model}.
    The prior distribution is chiral EFT-agnostic.
    For comparison, the model parameters used in Ref.~\cite{Hebeler:2013nza} are:
    $\alpha=5.87$, $\eta=3.81$ (point estimates) and $\alpha_L=1.38$, $\eta_L=0.87$, where we give for the latter two only the mean values of the underlying distributions. %
    }
    \label{fig:meta_model_prior_params}
\end{figure*}

\begin{figure*}[tb]
    \centering
    \includegraphics[width=\textwidth]{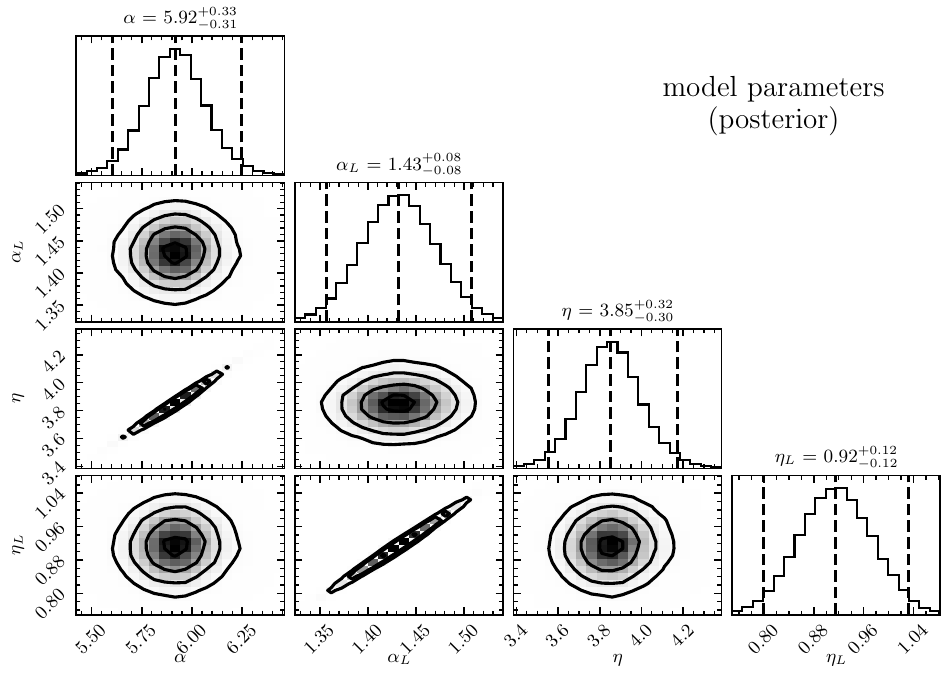}
    \caption{%
    Same as in Fig.~\ref{fig:meta_model_prior_params} but for the chiral EFT-informed posterior distribution of the model parameters.
    Note that $\alpha$ and $\eta$ match those of the prior distribution in Fig.~\ref{fig:meta_model_prior_params} because these parameters do not contribute to PNM and the model calibration is only informed by PNM data, as discussed in the main text. %
    }
    \label{fig:meta_model_posterior_params}
\end{figure*}

Figures~\ref{fig:meta_model_prior_params} and~\ref{fig:meta_model_posterior_params} show the prior and posterior distribution functions for the parameters of the model~\eqref{eq:eos_model}, respectively. 
The physics-informed prior distribution is chiral EFT-agnostic and only informed by empirical ranges for the saturation point, symmetry energy, and its slope parameter, as discussed in Sec.~\ref{sec:eos_modeling_framework}. 
For comparison, the model parameters used in Ref.~\cite{Hebeler:2013nza} to describe the SNM EOS are the following point estimates: $\alpha=5.87$ and $\eta=3.81$.
Our prior distribution is consistent with these point estimates but explores a larger parameter space because it was fit to the empirical saturation distribution~\eqref{eq:emp_sat_point}.
By comparing Figs.~\ref{fig:meta_model_prior_params} and \ref{fig:meta_model_posterior_params}, one observes that the uncertainties in $\alpha_L$ and $\eta_L$ are significantly reduced at the posterior level relative to the prior, whereas those in $\alpha$ and $\eta$ remain (up to statistical fluctuations) unchanged. 
This behavior is expected because the model is deliberately calibrated only to PNM data, while $\alpha$ and $\eta$ govern the SNM EOS and are thus constrained only by the prior (see also Sec.~\ref{sec:eos_modeling_framework}).

\begin{figure}[tb]
    \centering
    \includegraphics[width=\columnwidth]{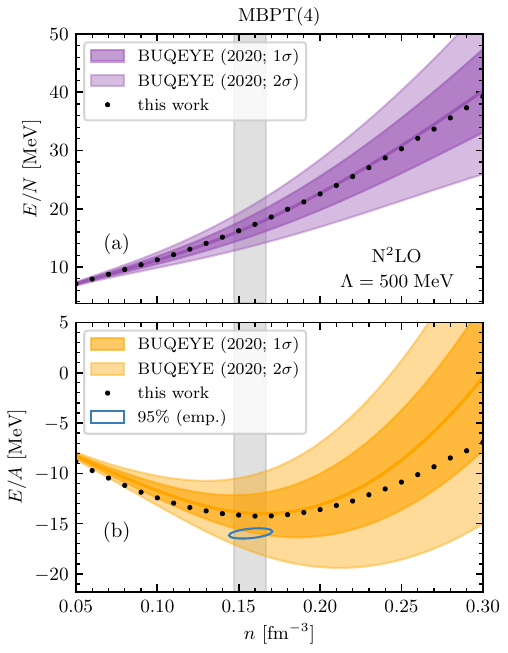}
    \caption{%
    Same as Fig.~\ref{fig:eos_buqeye_comparison}, but for the \NNLO interaction.%
    }
    \label{fig:eos_buqeye_comparison_n2l0}
\end{figure}

Figure~\ref{fig:eos_buqeye_comparison_n2l0} shows our MBPT(4) results for PNM [Fig.~\ref{fig:eos_buqeye_comparison_n2l0}(a)] and SNM [Fig.~\ref{fig:eos_buqeye_comparison_n2l0}(b)] using the DHS potentials, analogous to Fig.~\ref{fig:eos_buqeye_comparison}, but for the \NNLO interaction. 
Our findings at \NNLO and \NNNLO are qualitatively similar. 
In PNM across all densities shown, and in SNM at $n \lesssim 1.5 n_0$, normal-ordered 3N contributions at fourth order and residual 3N contributions at third order in the MBPT expansion have a negligible impact on the EOS. 
However, in SNM at $n \gtrsim 1.5 n_0$, these additional contributions, which were not considered in Refs.~\cite{Drischler:2017wtt,Drischler:2020yad,Drischler:2020hwi}, markedly alter the density dependence of the EOS, leading to increased binding. 
Nevertheless, our results remain well within the $1\sigma$ uncertainty band of the EFT truncation error analysis reported in Refs.~\cite{Drischler:2020yad,Drischler:2020hwi}.
However, one should note that the EFT truncation error is roughly a factor of two larger at \NNLO than at \NNNLO. 
For example, Refs.~\cite{Drischler:2020yad,Drischler:2020hwi} determined $E(n=n_0^\star,\delta=1)/A \approx 17.3 \pm 1.8 \MeV$ at \NNLO and $16.4 \pm 1.0 \MeV$ at \NNNLO, as well as $E(n=n_0^\star,\delta=0)/A \approx -14.0 \pm 1.9 \MeV$ at \NNLO and $-14.1 \pm 0.9 \MeV$ at \NNNLO.
See Refs.~\cite{Drischler:2020yad,Drischler:2020hwi} for more details.

\section{Spin-Isospin Traces} 
\label{app:traces}

As discussed in Sec.~\ref{sec:mbpt_diagrams}, the evaluation of each diagram involves integrating over momenta and summing over isospin and spin degrees of freedom for propagators. 
Diagrams in each additional MBPT order include a different normal ordered two-body vertex with two different propagators, each carrying isospin and spin degrees of freedom.
In earlier calculations~\cite{Drischler:2017wtt}, the integrand function evaluated each vertex at the given momenta for every spin and isospin configuration, yielding $2^4 \times 2^4 = 256$ matrix elements per vertex. 
A spin-isospin trace is then performed over the isospin and spin configurations for all propagators in the diagram.  
At fifth order, for example, there are 10 spin and 10 isospin degrees of freedom.  
This exponential growth in trace cost with MBPT order accounts for a large fraction of the integrand evaluation time at fifth order and becomes dominant at sixth order. 

In our improved implementation, we have derived, for each diagram's table entry, the subset of isospin configurations consistent with isospin conservation at each vertex. 
This approach yields a maximum count of 36 and 72 configurations across all diagrams at fifth and sixth order, respectively, a significant improvement over $2^{10}$ and $2^{12}$, respectively.

For further optimization, we organize the trace calculation into an outer loop over valid isospin configurations and an inner loop structure over spins.   
Outside the inner spin loop, the 256-entry vertex matrix element tables are subsetted to the isospin configuration under evaluation, leaving 16 matrix elements to be indexed by the 4 spins of the connected propagators, thereby saving time in the inner loops.
 
The inner spin summation is then reorganized by finding the most connected vertex pair.   
The spins of the propagators with at least one connection outside the pair are iterated over in an outer spin loop.  
All calculations about other vertices and propagators are evaluated outside the inner spin loop over the pair propagators.  
The inner spin loop is unrolled with length 2 (i.e., first found at fifth order), 4, or 8, depending on the number of connections in the pair to complete the contribution to the trace. 
The net effect is to use analysis of the diagram structure to hoist expensive calculations from inner loops to outer loops, yielding a significant speedup.

We have verified the above optimizations by cross-checking against an unoptimized trace. 
The result of these additional optimizations is that the spin-isospin trace is once again an insignificant portion of the integrand evaluation, even at high order.

\section{Computational scaling} 
\label{app:comp_scaling}

\begin{figure}[b]
    \begin{center}
    \includegraphics[width=\linewidth]{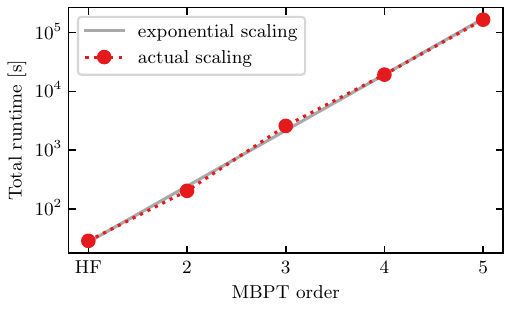}
    \end{center}
    \caption{%
    Exemplary runtimes of the developed MBPT framework when applied to the Hebeler~\etal\ interaction ``$(2.0/2.0)$'' at $n = 0.16 \fmiq$ in SNM.
    The dotted red line shows the total runtime $t_\mathrm{run}$ (i.e., the number of nodes multiplied by the runtime per node on Summit) to compute each MBPT order, not including residual diagrams.  
    The observed runtimes increase exponentially, approximately by a factor of 10 per MBPT order.
    See the main text for details.%
    }
    \label{fig:orderruntime}
\end{figure}

Figure~\ref{fig:orderruntime} illustrates the computational scaling of the developed MBPT framework up to fifth order when applied to the Hebeler~\etal\ interaction ``$(2.0/2.0)$'' at $n = 0.16 \fmiq$ in SNM on Summit. 
We choose SNM here because the MBPT calculations are more demanding than, e.g., those in PNM.
The dotted red line represents the total runtime $t_{\mathrm{run}}$, which is calculated by multiplying the number of nodes by the runtime per node for each MBPT order. 
Additionally, the gray line displays the results of a least-squares fit, resulting in $t_{\mathrm{run}}(k) \approx 27.6 \times 8.9^k \, \text{s}$, where $k$ denotes the MBPT order and HF corresponds to $k=0$.
The runtime increases roughly by an order of magnitude for each successive order, so (although substantial) lower than one might have expected, considering the significant increase in the number of diagrams at each MBPT order and the increased dimensionality of the momentum integrals (see Table~\ref{tab:mbpt}).
Residual diagrams have been omitted from this comparison because they are only implemented at second and third order, and the integrands have 3N interaction nodes with varying runtime characteristics.

The runtimes depicted in Fig.~\ref{fig:orderruntime} are affected by various factors. 
We observe that many diagram groups, with their internal cancellations, produce shrinking total results and errors at high MBPT orders. 
They converge to the error tolerance as the number of samples decreases.
However, the dimensionality of the integrations and the number of diagrams to be evaluated increase rapidly with order, thereby increasing the runtime per sample. 
Overall, the net effect of these competing factors is an approximately tenfold increase in runtime for each successive order.

\begin{figure}[tb]
    \begin{center}
    \includegraphics[width=\linewidth]{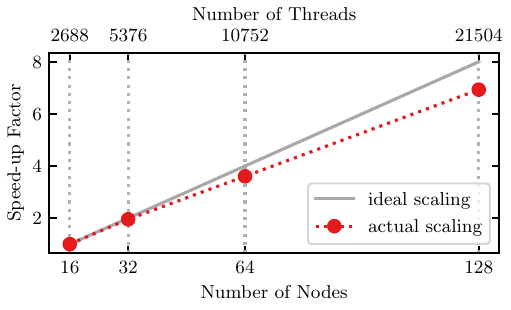}
    \end{center}
    \caption{%
    Exemplary speed-up factors of our MBPT framework when applied to the Hebeler~\etal\ interaction ``$(2.0/2.0$)'' at $n = 0.28 \fmiq$ in SNM.
    The runs evaluate 142 diagrams (corresponding to one diagram group) at fifth order with dimensionality $27=N_\text{dim}-3$, reduced by 3 due to rotational symmetries; see Table~\ref{tab:mbpt}.  
    The dotted red line shows the speed-up factors of our framework as a function of the number of compute nodes (and threads) when evaluating a set of 142 diagrams (i.e., one diagram group) at fifth order. 
    The speed-ups are normalized to the run on 16 nodes (2688 threads), and the connecting lines are included to guide the eye. 
    The solid gray line indicates the ideal scaling. For the 128-node run, the ideal speed-up relative to 16 nodes is $128/16=8$.
    The measured speed-up is 6.9, so close to the ideal value.%
    }
    \label{fig:pvegas_scaling}
\end{figure}

Figure~\ref{fig:pvegas_scaling} illustrates how the highest-order SNM calculations scale with the number of nodes.
As in Fig.~\ref{fig:orderruntime}, the calculations are based on the Hebeler~\etal\ interaction ``2.0/2.0.'' 
They were run on Summit with 28 OpenMP threads per rank, 6 ranks per node, and 1 GPU per rank.  
OpenMP threads in a rank shared a single GPU via CUDA streams.
The actual scaling (depicted by the red points and dashed lines) is nearly linear because \pvegas requires only a small amount of data (i.e., partial sums) to be exchanged between the MPI ranks.
The latency of MPI operations increases slowly with the number of nodes, and the computation of new piecewise-linear maps for importance sampling is the only operation not distributed across nodes but instead distributed across threads on rank 0, resulting in a slow roll-off from the ideal line.   
We leverage this scaling, in combination with \mpijm, to improve the efficiency of running many integrations.   
By increasing the node count for challenging integrations, we lower the probability that they will still be running at the end of a supercomputer batch run, thereby reducing wasted compute time.

\bibliographystyle{apsrev4-2}
\bibliography{bayesian_refs,bib}

\end{document}

%% file: buqeye_macros.tex
\makeatletter
\newcommand\newsubcommand[3]{\newcommand#1{#2\sc@sub{#3}}}
\def\sc@sub#1{\def\sc@thesub{#1}\@ifnextchar_{\sc@mergesubs}{_{\sc@thesub}}}
\def\sc@mergesubs_#1{_{\sc@thesub#1}}

\newcommand\newsupcommand[3]{\newcommand#1{#2\sc@sup{#3}}}
\def\sc@sup#1{\def\sc@thesup{#1}\@ifnextchar^{\sc@mergesups}{^{\sc@thesup}}}
\def\sc@mergesups^#1{^{\sc@thesup#1}}
\makeatother

\DeclareMathAlphabet{\mathbcal}{OMS}{cmsy}{b}{n}

\newcommand{\etal}{\textit{et~al.}\xspace}

\newcommand{\fmi}{\, \text{fm}^{-1}}

\newcommand{\fmiq}{\, \text{fm}^{-3}}
\newcommand{\keV}{\, \text{keV}}
\newcommand{\MeV}{\, \text{MeV}}

\newcommand{\NNLO}{\ensuremath{{\rm N}{}^2{\rm LO}}\xspace}
\newcommand{\NNNLO}{\ensuremath{{\rm N}{}^3{\rm LO}}\xspace}

\newcommand{\NkLO}[1]{\ensuremath{\mathrm{N}^{#1}\mathrm{LO}}\xspace}

\newcommand{\ordervec}{\vec}

\newcommand{\inputvec}{\mathbf}

\newsubcommand{\ckvec}{\ordervec{c}}{k}

\newsubcommand{\bkvec}{\ordervec{b}}{k}

\newsubcommand{\ckvecset}{\ordervec{\inputvec{c}}}{k}

\newsubcommand{\ckvecapprox}{\mathbf{c}'}{k}
\newsubcommand{\ckvecapproxset}{\mathbf{C}'}{k}

\newsubcommand{\bkvecapprox}{\mathbf{b}'}{k}
\newsubcommand{\bkvecset}{\mathbf{B}}{k}
\newsubcommand{\bkvecapproxset}{\mathbf{B}'}{k}

\newcommand{\genobs}{y}

\newsubcommand{\genobsvec}{\ordervec{\genobs}}{k}
\newsubcommand{\genobsvecset}{\ordervec{\inputvec{\genobs}}}{k}

\newsubcommand{\akvec}{\mathbf{a}}{k}

\newsubcommand{\akvecapprox}{\mathbf{a}'}{k}
\newsubcommand{\akvecset}{\mathbf{A}}{k}
\newsubcommand{\akvecapproxset}{\mathbf{A}'}{k}

{}  

\DeclareMathOperator{\pr}{pr} 
\newcommand{\given}{\,|\,}  

\newcommand{\normal}{\mathcal{N}}

\def\diffd{\mathrm{d}}  

\DeclareDocumentCommand\differential{ o g d() }{ 
    \IfNoValueTF{#2}{
        \IfNoValueTF{#3}
            {\diffd\IfNoValueTF{#1}{}{^{#1}}}
            {\mathinner{\diffd\IfNoValueTF{#1}{}{^{#1}}\argopen(#3\argclose)}}
        }
        {\mathinner{\diffd\IfNoValueTF{#1}{}{^{#1}}#2} \IfNoValueTF{#3}{}{(#3)}}
    }
\DeclareDocumentCommand\dd{}{\differential} 

\newcommand{\pathd}{\mathcal{D}}  

\DeclareDocumentCommand\pathdifferential{ o g d() }{ 
    \IfNoValueTF{#2}{
        \IfNoValueTF{#3}
            {\pathd\IfNoValueTF{#1}{}{^{#1}}}
            {\mathinner{\pathd\IfNoValueTF{#1}{}{^{#1}}\argopen(#3\argclose)}}
        }
        {\mathinner{\pathd\IfNoValueTF{#1}{}{^{#1}}#2} \IfNoValueTF{#3}{}{(#3)}}
    }

%% file: macros.tex
\usepackage{xifthen}

\newcommand{\mpijm}{\texttt{Mpi\_Jm}\xspace}
\newcommand{\pvegas}{\texttt{PVegas}\xspace}

\newcommand{\Vmed}[1]{\bar{V}_{\mathrm{eff}}^{#1}}

\newcommand{\CC}{C\nolinebreak\hspace{-.05em}\raisebox{.4ex}{\tiny\bf +}\nolinebreak\hspace{-.10em}\raisebox{.4ex}{\tiny\bf +}\xspace}

\newcommand{\iso}[2]{#1_#2}

\newcommand{\kF}[1][]{%
  \ifthenelse{\isempty{#1}}%
    {k_\text{F}}
    {k_\text{F}^{(#1)}}
}

\newcommand{\nk}[1][]{%
  \ifthenelse{\isempty{#1}}%
    {n_{k}^{(\tau)}}
    {n_{k_#1}^{(\tau_#1)}}
}

\newcommand{\eps}[1][]{%
  \ifthenelse{\isempty{#1}}%
    {\iso{\varepsilon}{{\tau}}(k)}
    {\iso{\varepsilon}{{\tau_#1}}(k_#1)}
}

\newcommand{\rev}[1]{#1}